\documentclass[preprint]{aastex}

\def\edcomment#1{\iffalse\marginpar{\raggedright\sl#1\/}\else\relax\fi}
\begin{document}
\title{A Multiwavelength Study of Stephan's Quintet}
\author{Jack W. Sulentic}
\affil{Physics \& Astronomy, University of Alabama, USA}
\author{Margarita Rosado \& Deborah Dultzin-Hacyan}
\affil{Instituto de Astronomia, UNAM, Mexico}
\author{Lourdes Verdes-Montenegro}
\affil{IAA, Granada, Spain}
\author{Ginevra Trinchieri}
\affil{OAB, Milan, Italy}
\author{Cong Xu}
\affil{IPAC, Caltech, USA}
\author{Wolfgang Pietsch}
\affil{MPE, Garching, Germany}

\begin{abstract}
Stephan's Quintet (SQ) is a compact group that we find in an atypical
moment when a high velocity intruder is passing through it. The
intrusion is particularly interesting because a previous intruder had
stripped most of the gas from the group members.  This debris field was
shocked in the ongoing collision with the new intruder.  This
evolutionary history agrees well with observations and explains how a
strongly interacting system can show low levels of star formation.  We
present new multiwavelength data including previously unpublished ROSAT
X-ray, H$\alpha$ interference filter/Fabry-Perot, ISO MIR/FIR and radio
line and continuum images. These observations and previously published
data provide new insights as well as support for some previous
hypotheses.  1) Fabry-Perot and HI velocities allow us to unambiguously
distinguish between gas associated with SQ and the new intruder. 
2) Most detected emission regions are found in the remnant
ISM of the NI which allows us to infer its size and present physical
state.  3) The few emission regions associated with the stripped ISM of
SQ include the best candidate tidal dwarf galaxy. 4) Multiwavelength
data suggest that strong MIR/FIR emission from the Seyfert 2 nucleus of
NGC7319 comes from dust heated directly by a power-law continuum rather
than a starburst. 5) The correspondance between extended X-ray/radio
continuum/forbidden optical emission confirms the existence of a large
scale shock in SQ.  6) We confirm the presence of two stripped spiral
members in the process of transformation into E/S0 morphology.
Finally, 7) Observations are consistent with the idea that the
collision in SQ is ongoing with possible detection of HII region
ablation and Rayleigh-Taylor instabilities.
\end{abstract}


\section{Introduction}

Compact groups are aggregates of four or more galaxies showing
projected separations on the order of $\sim$30-40 kpc (assuming
H$_o$=75 km s$^{-1}$ Mpc$^{-1}$) which imply space densities similar to
the cores of rich clusters. Compact groups are high density
fluctuations usually located in non-cluster environments (Sulentic
1987; Rood \& Williams 1989).  Their importance is twofold:  1) they
are ideal laboratories for studying the effects of extreme galaxy
interactions and 2) they are low redshift analogs of processes believed
to be very important at high redshift.  One can study the groups either
statistically or individually.  The former approach still suffers from
effects of sample selection bias and incompleteness.  For example, only
60 of the 100 groups cataloged by Hickson (1982) actually satisfy the
initial selection criteria (Sulentic 1997).  The remaining objects
either violate the selection criteria, extend them beyond the
originally stated limits or involve triplets (with a fourth discordant
redshift galaxy projected).  Since it is unclear whether triplets share
the same properties as N$>$3 accordant redshift systems (Sulentic
2000), it appears safer at present to treat them separately. In any
case, if they represent groups in the process of formation by
sequential acquisition of neighbors,  they will not show the same level
of interaction induced phenomena as richer systems.  We note that 16
triplets included in the compact group sample studied by
Verdes-Montenegro et al. (2001) did not show the significant hydrogen
deficiency found for systems with four or more members.  They also show
larger velocity dispersions than n$\geq$4 systems (Sulentic 2000)
suggesting that they are unbound systems. A new southern hemisphere
sample of compact groups (Prandoni et al. 1994; Iovino 2000),
selected with automated techniques, promises to minimize and quantify
effect of bias, when a suitable multiwavelength database better defines
its properties.

Stephan's Quintet (SQ) is the ideal candidate for detailed study
because it is bright and because it is in a rare but important stage of
dynamical evolution.  Inferring group properties from SQ is not
unreasonable because it is typical of the compact group phenomenon as
defined in the Hickson (1982) catalog (see also Prandoni et al. 1994)
only showing more spectacular properties because of an ongoing
collision. The best example of an ``active'' compact group in the
southern hemisphere may involve the Cartwheel galaxy (Wolter et al. The
crossing time for such collisions is so short (few $\times$ 10$^7$
years) that more than one ``SQ'' is unlikely to be found in a sample of
one hundred compact groups.  The frequency of such collisions is a
direct function of local galaxy density and SQ is not in an unusually
dense environment (Sulentic 1987).  1999) where the most recent
collision occurred a few $\times$10$^8$ years ago. The distribution of
component velocities relative to the first ranked member in Hickson
groups (Sulentic 2000) suggests that SQ is not the only group with 
a possible  high velocity intruder nearby.

A ``two intruder'' (old intruder=NGC7320c=OI; new intruder=NGC7318b=NI)
evolutionary model has been proposed for SQ (Moles et al.1997) that
may be relevant to most compact groups and that may have relevance
for our understanding of interactions at high redshift. This model
forms the basis for our interpretations of the multiwavelength data.
In section 2 we discuss the new observations and their reductions as
well as data harvested from archival sources.  In section 3 we discuss
past work on SQ and compact groups in general. We combine new and old
observations in Section 4 to show how they are consistent with the two
intruder scenario. Section 5 summarizes important results and their
implications.

\section{New and Harvested Multiwavelength Observations}

We present new observations at X-ray, optical, infrared and radio
wavelengths as well as optical archival data that include the highest 
resolution (HST) and highest sensitivity (CFHT)  optical images. 

\subsection{ROSAT X-ray}

We obtained new ROSAT HRI  (Tr\"umper 1983, Pfeffermann et al. 1987)
data in Dec 1996-Jan 1997 with a total observing time of 65 ksec split
into 35 intervals (OBIs).  These observations followed a shorter one
($\sim$ 23 ksec), already discussed (Pietsch et al. 1997), which we
also include in the present analysis.  We excised time patches of data
shorter than 40 sec. (caused by high background rejection in the
standard analysis and/or high voltage excursions, for a total of 600
sec.) and lowered the tolerance to low count rates, that excludes a
higher percentage of high particle background data, mostly in proximity
to radiation belts.  This screening resulted in a loss of 6.4 ksec of
data, but should ensure a very clean dataset. The same cleaning was
applied to the old dataset, so the total observing time on SQ is
$\sim$77.5k sec.

Due to the lack of strong point sources in the HRI field of view,  we
could not improve on the short term aspect solution used by the
Standard Analysis Software System (SASS, Voges et al. 1992).  We
however verified that no major time-dependent effects are present in
the data by accumulating and comparing images in three $\sim$ 20 ksec
intervals.  We found that:  a) the positions of the few field sources
were the same and b) a possible residual systematic effect of improper
de-wobbling might still be present.  Further checks confirmed a
residual wobbling-related effect, so that no structure on scales of
$7''-10''$ could be trusted.  We also evaluated the absolute pointing
of the observations by searching for optical counterparts to all of the
X-ray detections in the field of SQ.  None of the sources outside of SQ
coincide with known objects ($e.g$.  from SIMBAD or NED) but faint
counterparts are visible on the DSS2 for three sources.  We re-aligned
the absolute coordinates of the X-ray map on these sources, using in
addition the radio position of the Seyfert nucleus (van der Hulst \&
Rots 1981).  This produced a $0.4^s$ shift to the E.  Our final
positions should be accurate to $2''-3''$.

We considered merging both HRI datasets in order to improve the
statistical significance of our results.  However, since the source is
not at the same detector position in both pointings (it was $\sim 4'$
off-axis in the first observation), and in light of known spatial
inhomogeneities in detector gain, which also changes with
time\footnote{ see the discussion in the The ROSAT HRI Calibration
Report and Users' Handbook, available on-line at {\tt
http://hea-www.harvard.edu/rosat/}}, we decided against it, and we use
the datasets independently.  Most of the following discussion is based
on the second and longer observation with the shorter one used for
consistency checks.  To limit the changes due to gain effects, while
improving on the signal-to-noise ratio, we have selected data in PHA
channels 1 to 10 for both observations, which considerably reduces the
background contribution (see also the on-line documentation; note that
this is  a larger range than used in  Pietsch et al.  1997, where only
PHA channels 4-8 were included).  The background was obtained from a
$4'-6'$ annulus around the field center in both datasets.  In the first
observation, since the source itself is at $\sim 4'$ off-axis and is
therefore included in this annulus, a circle of $3'$ radius centered on
SQ is masked out from the background region.

Figure 2 (left panels) shows two X-ray images of SQ involving: a)
adaptively smoothed (upper) and c) gaussian filtered (lower) data.  The
adaptive smoothing algorithm provided with the CIAO software was used
(see {\tt http://asc.harvard.edu/} for further details). We have used
the FFT method with 2$\sigma$ minimum significance.  Table 1 summarizes
the contribution of each feature, to the old and the new observations.
The flux is obtained from the total count rate, converted assuming a
$raymond$ spectrum with kT=1 keV and the line-of-sight galactic
absorption N$\rm _H$=7$\times 10^{20}$ cm$^{-2}$.  The flux
determination is almost independent of the choice of a spectral model,
given the relatively narrow energy range of ROSAT, for reasonable
values of the absorbing column density (see Pietsch et al. 1998). This
assumption will not be correct if significant absorption is present,
e.g. in the Seyfert 2 nucleus.  We have defined 4 regions: 1) a circle
at the position of the Seyfert galaxy NGC7319, with r = 18$''$; 2) a
circle at the position of NGC 7318a (r=15$''$); 3) a rectangle of $0.9'
\times 1.6'$, centered on the elongated NS feature; and 4) a smaller
rectangle ($0.4'\times \sim 0.8'$) that includes only the higher
surface brightness part of this feature.

Figure 2 (right panels) show: 2b) 21cm radio continuum (upper) and 2d)
forbidden [NII]$\lambda$6583 (lower) images. The radio image (see
Section 2.5) shows the same structure as the X-ray in: a) the elongated
feature (interpreted as a collisional shock), b) the extension of this
feature to the extreme NW of SQ and c) the detection of the elliptical
galaxy member NGC7318a. The radio continuum image of the extended
(nonthermal shock) feature is relatively stronger and broader on the
south edge which may reflect attenuation of the X-ray emission due to
the larger hydrogen column of foreground NGC7320 which extends over
this region. The peak HI column in NGC7320 N$_H$= 1.6$\times$10$^{21}$
cm$^{-2}$ is slightly more than twice galactic  and the distribution of
HI is flat so that there is more than double galactic column
(9.0$\times$10$^{20}$ cm$^{-2}$ from NGC7320 and 7.0$\times$10$^{20}$
cm$^{-2}$ from the Milky Way) in the region where the X-ray emission
from the shock is weakest relative to radio continuum.

We attempted to quantify the significance of any low surface brightness
component that might be present in SQ  by measuring emission in several
features visible on the X-ray maps.  Figure 3 shows the azimuthally
averaged profile of the total emission as well as the emission averaged
in two opposite quadrants along the NS and EW directions.  We derived
the profiles in concentric annuli of increasing size, centered on the
peak of the emission in the extended feature as seen in the adaptively
smoothed map (Figure 2a).  The azimuthally averaged profile shows a
clear excess over the background out to r$\sim 1.5'$.  The emission
appears to be extended roughly equally in the NS and in the EW
directions.  Some very low surface brightness emission could extend to
r$\sim 3'$, mostly in the EW direction (right panel in Figure 3).
 Recently obtained CHANDRA data confirm the existance of such emission
(Trinchieri et al. in preparation).

\subsection{H$\alpha$ Interference Filter Imaging}

Interference filter observations were obtained in August 1997 with the
2.2m telescope at Observatorio Astronomico Hispano Aleman on Calar
Alto, Spain (see also Xu et al. 1999). They are not the first emission
line images obtained for SQ (Arp 1973; Moles et al. 1997; Vilchez \&
Iglesias-Paramo 1998; Severgnini et al. 1999) but they are the most
sensitive and cover the entire group and effectively separate group
from NI emission.  Two narrowband filters (667/7 and 674/8) centered at
6667 (FWHM=66$\AA$) and 6737$\AA$ (FWHM= 76$\AA$) allowed us to
separate most of the SQ emission (6300-7000 km/s) from that of the NI
(5600-6000 km/s). The H$\alpha$ transmission for the 667/7 filter was
0.49 for the 5700 km/s emission and 0.04 for SQ emission while the
674/8 filter yielded a transmission of 0.49 for SQ and 0.06 for 5700
km/s emission. The region of strongest overlap involves emission from
the NI at $\sim$ 6000 km/s located north of the nucleus.  The continuum
was subtracted from both images using an R band image. Both maps were
smoothed with a 2 arcsec Gaussian beam in order to increase sensitivity
to diffuse emission (Xu et al.  1999). The SQ and NI emission line
images are shown in Figures 4a and 4b respectively, with Figures 5a and
5b presenting, respectively, the emission line images with contours for (VLA, $\S$2.5)
HI emission in the velocity ranges 6500-6800 and 5600-6000 km/s,
respectively, superimposed.  Figure 6 (bottom panel) shows the
appropriate wavelength ranges where emission from  H$\alpha$,
[NII]$\lambda$6583$\AA$ and the [SII] doublet $\lambda\lambda$6717,
6731$\AA$ are observed. In the lower panel we also indicate the filters
used with appropriate FWHM. The diagram allows one to assess the degree
of overlap for all the interference filter observations described in
this paper.

We also used the 2.1m telescope at Observatorio Astron\'omico Nacional
(at San Pedro M\'artir, Baja California, M\'exico) to obtain  three 20
minute exposures through a narrow band filter centered at  6731$\AA$
(FWHM=10$\rm{\AA}$). The observations were made with the Fabry-Perot
interferometer instrument (PUMA without an  etalon) in the direct
imaging mode.  The resultant images are sensitive only to
[NII]$\lambda$6583$\AA$ (+continuum) emission in the 6740-6970 km/s
velocity range (SQ ISM) and should therefore image diffuse [NII]
emission associated with the shock as well as discrete emission from
HII regions.  The full velocity range of this emission is estimated to
be 6300-7000 km/s based upon our Fabry-Perot measures.  This image is
shown in Figure 2d. Figure 6 indicates that we simultaneously image the
[SII]$\lambda$6717$\rm{\AA}$ emission in NGC7320 where several of the
highest density HII regions are enhanced.

\subsection{Fabry-Perot H$\alpha$ Imaging}

The observations were carried out during the nights of October 27, 28
and 30, 1997 and September 10, 1999 with the  Fabry-Perot
interferometer PUMA attached to the f/7.9 Ritchey-Chretien focus of
the 2.1 m telescope at the Observatorio Astron\'omico Nacional.  
The PUMA setup is composed of a scanning Fabry-Perot (F-P) interferometer, 
a focal reducer with an f/3.95 camera, a filter wheel, a calibration system
and a Tektronix 1K$\times$1K CCD detector (Rosado et al. 1995).  The main
characteristics of the instrumental setup are the following: CCD
readout was binned 2$\times$2, giving a pixel size of 1.2 arcsec
(equivalent to $\sim$500 pc at 90 kpc)  with a 10 $\times$ 10 arcmin 
field of view.

The F-P interferometer has an interference order of 330 at 6563 \AA\ 
and its free spectral range of 934 km/s was scanned in 48 steps with a
sampling resolution of 19 km/s, yielding a data cube of 48 $\times$
512  $\times$ 512 elements. F-P observations of SQ are complicated
because of the wide range of velocities present in three overlapping
ranges (see Figure 6) and because the emission is so complex and
extended. The large field of view and free spectral range of PUMA were
indispensable for interpreting the F-P data. We obtained data cubes in
the lines of H$\alpha$ in the velocity range of the NI and
[NII]($\lambda$ 6583 \AA) in the velocity range of SQ.  No filter was
available for a zero order observation of H$\alpha$ but it
appears at second order in our velocity cubes with almost identical
channel--velocity correspondance (the velocity separation of the two
lines is almost identical to the free spectral range of PUMA). During
our first observing run we obtained two data cubes at 6750\AA, three at
6720 \AA\ and two at 6680 \AA, each one with an exposure time of 60s
per channel.  During the second run we obtained three data cubes at
6680 \AA, each with  120 s integration time per channel. Similar
quality data cubes were co-added in order to enhance the S/N ratio. We
thus obtained total exposure times of 1.6, 2.4 and 1.6 hours for the
observations at 6750, 6720 and 6680 \AA,  in the first observing run,
respectively,  and a total exposure time of 4.8 hours for the
observations at 6680 \AA\ in the second observing run. The $\pm$1
orders fall  at the edges (half power points) of the filter ranges
shown in Figure 6.

The phase calibration was made by taking data cubes of a calibration
lamp before and after the SQ exposures.  When the exposure
times of the object data cubes were of more than two hours, we obtained
an additional calibration data cube between object exposures in order
to ensure that no instrument flexure had occurred.  We  used a neon
line at 6717.04 $\rm{\AA}$\ for calibrating the object cubes. Reductions were
carried out using the software package CIGALE (Le Coarer et al. 1993).
Our data cubes in H$\alpha$ are contaminated with night sky line
emission which was subtracted using an interactive routine in CIGALE.
PUMA data for SQ and the NI are shown as contour maps in
Figure 7ab. Velocities were obtained by PUMA for virtually all emission
features seen in the Calar Alto IF images (Figure 4ab) except for
velocity smeared emission in the shock. Table 2 presents the velocities
(column 2) for emission regions identified  by number (column 1) in
Figure 7.  The left listing represents H$\alpha$ in the NI
(Figure 7b) while the right gives values for [NII] in SQ (Figure 7a).
The left listing also presents literature measures for comparison (CO
and HI measures are presented in the text). H$\alpha$ emission with SQ
velocities in the range 6350-6414 km/s is also detected in 18, 19 and
34. Column 3 lists velocities from Plana et al.  1999 whenever one of
their possible velocities fell within  $\pm$100 km/s of our adopted
value. Other literature values are given in Column 4 (h denotes
Gallagher et al.  (2001); f denotes Arp (1973); o denotes corrected
value for Ohyama et al. (1998) which is contaminated by broad emission
at SQ velocities; all others are from Moles et al. (1998). Data quality
is indicated by *, + or - following a velocity which correspond to
1$\sigma$ uncertainties of 10, 20 and 30 km/s respectively. The latter
usually correspond to regions where shock velocity smearing is seen.
Fewer SQ velocities were obtained because most emission in that
velocity range is in/near the shock where smearing and multiple
velocity overlap  often make reliable measurement uncertain or
impossible.

All of our derived velocities in the NI  and SQ that overlap
with HI emission clouds (see Figures 5ab) show very close agreement
with the HI derived velocities (HI velocities and channel maps are
given in Williams et al. 2001). We note that a number of our velocities
are in disagreement with those adopted by Plana et al. (1999). There
are two reasons for this: 1) the much smaller free spectral range of
the two Fabry-Perot etalons that they employed led to three or four
fold ambiguity about the correct source velocities and 2) many/most of
the emission regions  south of NGC7318ab have velocities outside of the
range of their interference filters.  Similarly the apparent
disagreement between our measures of two emission regions near or in
common with Ohyama et al.(1998) result from their interpreting the
strongest narrow line component as H$\alpha$ when it is, in fact
[NII]$\lambda$6583. The situation for the SQ velocity range is much
more complex.  All discrete source detections appear to correspond with
HII regions imaged in Figure 4a. Unlike the  NI data however,
the SQ velocities and distribution of emission features shows no
concentration or hint of ordered motion. Emission features in the far
NE have been discussed in connection with the tidal tails produced by
the OI. A velocity for residual H$\alpha$ (actually
[NII]$\lambda$6583 was measured) emission in (or projected on) NGC7319
was also given earlier. The remaining  emission more directly involved
with the NI will be discussed in section 4.2.

\subsection{ISO Imaging}

ISOCAM observations at 11.4 and 15 $\mu m$ were obtained on 1996 May 23
with the {\it Infrared Space Observatory}. The 15$\mu m$ observations
were previously published in Xu, Sulentic \& Tuffs (1999) where details
of the ISOCAM data reduction procedures can be found.  We briefly
summarize some key parameters that differ from the 15$\mu m$
observations. The 11.4$\mu m$ image was obtained using the ISOCAM
Long-Wavelength-Channel array (32$\times 32$ pixels) with the LW8
filter ($\lambda_0=11.4\mu m$, $\delta\lambda=1.3\mu m$), which is
sensitive to the  unidentified broad band emission feature (UIB) at
11.3$\mu m$.  Raster scans were made with $PFOV= 6''$, $M=12$ steps,
$\delta M= 48''$ (in-scan) and $N=20$ steps, $\delta N=6''$
(cross-scan). The basic data reduction was done using the CAM
Interactive Analysis (CIA) software\footnote{CIA is a joint development
by ESA Astrophysics Division and the ISOCAM Consortium led by the
ISOCAM PI, C. Cesarsky, Direction des Sciences de la Matiere, C.E.A.,
France.}. The effective angular resolution of the image is 10 arcsec
(FWHM).

ISOPHT observations at 60 $\mu m$ were obtained using the C100 camera
(3$\times 3$ pixels) on board ISO. The observations were carried out on
May 23, 1996, using the ISOPHT oversampling mapping mode (P32) with an
oversampling factor of 3.  The map has 16 scan lines and 9 pointings
per map line, covering about 13$'$ (in scan) $\times$ 6$'$ (cross scan)
on the sky. The 60$\mu m$ map was reduced using the newest P32 data
reduction package (Tuffs et al.  2001). This software takes advantage
of the high sampling rate and high redundancy of P32 maps.  The effect
of detector transients were carefully corrected. The angular resolution
of the 60$\mu m$ map is on the order of 30$''$ which is $\sim$ 50\%
better than the detector pixel size (45$''$). Figure 8a presents the
11.4 $\mu m$ image and Figure 8b the 60$\mu m$ image with 15 $\mu m$
contours superimposed.  The group is resolved even at 60$\mu m$ where
NGC7319 and 7320 are the strongest sources as with MIR wavelengths.
Table 2 presents 60$\mu m$ flux densities with estimated errors at
least 30\%. The total 60$\mu m$ flux is 1.26 Jy which is considerably
higher than S=0.89 Jy obtained by IRAS.

\subsection{VLA HI Mapping}

SQ was mapped in the velocity range 5511-6862 km/s with the VLA in the
C, CS and D configurations with subsequent data
reduction using AIPS. Details and complete sets of channel maps can be
found in Williams et al. (2001).  The integration ranges were
5575.4-5810.5, 5917.5-6088.8 and 6432.1-6776.1 km/s. The resultant
synthesized beam was 23.6$\times$17.4 arcsec with an rms noise in the
maps of 0.22mJy and a 3$\sigma$ HI flux limit of 0.015 Jy km s$^{-1}$
corresponding to an HI column density of 5$\times$10$^{19}$
atoms/cm$^2$.  The HI mass detection limit is
2.3$\times$10$^7$$M_\odot$/beam at the distance of SQ.  HI contours for
SQ (6500-7000 km/s) and the NI (5600-6000 km/s) are overlaid
on the corresponding H$\alpha$ interference filter images in Figures 5a
and 5b, respectively.

A radio continuum image was also constructed by averaging all line-free
channel maps. The effective radio continuum bandwidth was 2.05Mhz with
central frequency of 1416Mhz. The rms noise was 0.10mJy per beam. The
synthesized beamwidth is about three times larger that the highest
resolution continuum observations (van der Hulst \& Rots 1981) but has
about 3$\times$ higher  flux sensitivity. This map is shown in Figure
2b where comparable resolution adaptively filtered X-ray contours are
superimposed.

\subsection{CFHT Imaging}

Broadband B and R images were harvested from the 3.6m
Canada-France-Hawaii Telescope archive. The observations were obtained
on 21 August 1993 with the MOS/SIS (Multi-Object
Spectrograph/Subarcsecond Imaging Spectrograph) in direct imaging
mode.  Seeing was estimated to be 0.8 arcsec (Plana et al.  1999).
This archival data was kindly provided to us in reduced form by C.
Mendes de Oliveira. The data were published by Mendes de Oliveira et
al.  (2001) with a different emphasis. The B and R-band images
represent the average of 5 $\times$ 900 and 6 $\times$ 350 second
exposures, respectively. The B band image is shown in Figure 9c while a
B-R color map (see also Mendes de Oliveira et al. 2001) is shown in
Figure 9d.  CFHT provides the deepest images of  SQ.

\subsection{HST WFPC Imaging}

Broad-band WFPC2 B, V and I images for SQ were harvested from the
Hubble Space Telescope archive and processed using the standard HST
pipeline. Dithered sets of images were obtained on 30 December 1998 and
17 June 1999 with total exposure times for the B(F450W), V(F569W) and
I(F814W) datasets of 6700, 3200 and 2000 sec respectively. One WFPC2
pixel equals 0.09 arcsec ($\sim$35-40 pc). The earlier pointing was
centered on NGC7318ab and 7319 while the later included NGC7319 and the
full extent of the younger tidal tail. Rather than show the entire
field of view of the WFPC images we have inset in Figure 10 high
resolution vignettes from the B-band image of the most interesting
regions in SQ from this dataset.  The other utility of this data
involves B-V colors for various components of SQ. This data was
published by Gallagher et al. (2001) with a different emphasis. WFPC2
provides the highest resolution images of SQ.

\section{Previous Multiwavelength Results}

\subsection{Radio}
\subsubsection{Continuum Observations}

Radio continuum observations of systems as distant as SQ should involve
negligible thermal emission (e.g. foreground NGC7320 is not detected)
and should therefore be effective tracers of nonthermal  processes such
as AGN and shocks, both of which have often  been attributed to
galaxy-galaxy interactions. High resolution observations (Kaftan-Kassim
et al. 1975; van der Hulst \& Rots 1981) resolved  two continuum
sources in SQ from an apparently unrelated source at the north edge of
the group.  SQ detections involve:  1) the nucleus of the Seyfert 2
galaxy NGC7319 and 2) an extended source in the region  between NGC
7319 and the NI. The extended source  has been
interpreted  as: 1) the signature of an ongoing collision between
NGC7318a and 7319, 2) or between NGC7318a and b, as well as,  3) a
shock front involving a collision of the NI with something (Allen \&
Sullivan 1980; Shostak et al. 1984). In the former two scenarios the
enhanced radio continuum feature would arise from a very large number
of supernova remnants (Van der Hulst \& Rots 1981).  The point source
at the north end of the extended radio feature shows no correspondance
with emission at any other wavelength and is likely to be a background
source.

\subsubsection{Line Observations}

HI measures are a sensitive diagnostic of the degree of dynamical
perturbation in a galaxy or group of galaxies.  High resolution HI maps
of SQ (Shostak et al.  1984; Verdes-Montenegro et al. 2000) reveal that
almost all of the neutral hydrogen in SQ has been stripped from the
component galaxies. Molecular gas content is strongly correlated with
the star formation rate in a galaxy. High resolution observations of SQ
(Yun et al. 1997; Gao \& Xu 2000: \ Smith \& Struck 2001; Braine et al.
2001) indicate that CO emission is associated with: a) optically
identified dust clouds (inside, or superposed on,  NGC7319) and b)
tidal debris produced by the intruders.  While no measurable HI remains
in NGC7319, 3-4$\times$10$^9$ M$_{\odot}$ of M$H_2$ are found along
the same line of sight.

\subsection{Infrared}

IR emission is a sensitive measure of star formation activity in a
galaxy. This is especially true of MIR wavelengths where warm dust very
close to star forming regions is measured (Dultzin-Hacyan \& Masegosa
1990).  MIR radiation can be enhanced near sources of thermal (e.g.
star formation) or nonthermal (e.g. AGN) photons.  Deconvolved IRAS
(Verdes-Montenegro et al. 1998, Allam et al. 1996) and ISOCAM
observations (Xu et al. 1999 and Figure 8a) reveal that the bulk of the
MIR emission in SQ arises from : 1) the discordant redshift galaxy
NGC7320 (an Sd dwarf), 2) the nuclear region of NGC7319 and 3) two
detached  and compact starbursts, one in a tidal tail, produced by the
OI and the other in the debris field associated with the NI (both
involve strong CO detections; Smith \& Struck 2001, Braine et al.
2001).

\subsection{Optical}

\subsubsection{ Broad-Band Imaging Observations}

Broad band optical observations provide a direct measure of the
interaction morphology of a pair or group through detection of tidal
bridges/tails and diffuse light produced by dynamical stripping
processes.  Deep photographic images (Arp 1973) already revealed: 1)
the existence of two parallel tidal tails  and 2) an envelope of
diffuse starlight.  The former features extend towards the OI and
provide the best evidence that it has interacted with the group.  The
existence of two parallel tails suggest that it has been captured by SQ
unless a single passage can produce both features.  Surface photometry
measures suggest that both tidal tails are quite blue (Schombert et al.
1990).  An R-band measure of the luminous halo (Moles et al.  1998)
suggest that L$_{halo}$=L* implying a tidal stripping timescale of
about t$\sim$ 1Gyr assuming a reasonable rate of dynamical stripping
(see e.g.  Rabaca 1996). Two of the three core members of SQ (NGC 7317
and 7318a) show elliptical morphology with typical B-V colors$\sim$ 1.0
while NGC7319 shows barred spiral structure but without significant
evidence for an ISM (Zepf et al. 1991; Moles et al.  1998).

\subsubsection{Line Imaging, Photometry and Spectroscopy}

H$\alpha$ measures provide another direct measure of the star formation
rate. Forbidden [NII] $\lambda$6583 emission is usually included in
these measures and is more sensitive to shocks and nonthermal
processes.  Aside from the Seyfert nucleus, the other members of SQ
show little or no nuclear H$\alpha$ emission ((Vilchez \&
Iglesias-Paramo 1998ab; Iglesias-Paramo \& Vilchez 1999, 2001;
Severgnini et al. 1999). The strongest H$\alpha$ (and [NII]) emission
feature in SQ  lies between NGC7319 and the NI, apparently
associated with the event that gave rise to the radio synchrotron arc.

Very few published  high resolution and s/n spectroscopic observations
exist for SQ.  In part this is because the component galaxies lack
significant emitting gas and absorption line spectra for galaxies in
the range B=15-16 requires long integrations and/or large apertures.
This is especially true for the OI.  An HI 
recession velocity in RC3 for this galaxy is spurious  while a single
absorption line spectrum reveals V=6000$\pm$200 km/s implying a
velocity difference of $\sim$600 km/s relative to SQ (see $\S$4.1 for a
new value). Most published spectra involve HII regions near the radio
continuum/Halpha arc or in the NI which is the only SQ member
to retain significant gas. New Palomar 5m spectroscopy (Xu et al. 2001)
along the N-S H$\alpha$+[NII] feature in SQ finds most emission
consistent with shock excitation.

\subsection{X-ray}

X-ray observations of compact groups provide another independent
measure of high energy phenomena in the groups as well as information
about a hot IGM component. SQ was originally thought to have a  strong
diffuse X-ray component (e.g. from thermally heated gas; Sulentic et
al. 1996) but subsequent ROSAT HRI observations (Pietsch et al. 1997)
revealed that most of the emission comes from the Seyfert nucleus and
the region of the radio continuum and H$\alpha$ arc.  The X-ray
observations support the interpretation of this feature as a shock
interface between the SQ IGM and an NI infalling with a line of sight
velocity v=10$^3$ km/s.

\section{Discussion}

SQ is evolving through intergroup interactions but, especially through
the effects of sequential intrusions by neighbors from the associated
large scale structure.  The connection between ongoing and past events
is very clear; the SQ ISM  was produced/stripped  in one or more of the
past events involving an OI and part of it is now being
shocked by a collision with the NI.  The collision is unusual
because of the large velocity difference between SQ and the NI 
($\Delta$V$\sim$10$^3$ km/s). Figures 2, 4, 5, 8-10 show the new
multiwavelength data supplemented with harvested archival observations.
We amplify the evolutionary scenario advanced by Moles, Sulentic \&
Marquez (1997) by presenting old and new multiwavelength data that are
largely consistent with it.

 \subsection{SQ Past History: Old Intruder}

Past intrusions by the OI have significantly modified
SQ and created the stripped gaseous environment that gives rise to the
current shock and also contributed a significant mass of stripped stars
to the common halo of the group (Moles, Marquez \& Sulentic 1998).

SQ shows parallel tidal tails (Arp 1973; Arp \& Lorre 1976), with one
arcmin N-S separation ($\sim$ 25kpc assuming a distance of 90 Mpc),
extending towards the OI.  The existence of two tails suggests: 1)
that the OI has visited SQ at least twice and also that 2) it has
been captured. The latter inference is confirmed by the low line of
sight velocity difference between OI and SQ  ($\Delta$V$\sim$
0 km/s) obtained from our F-P observations (V=6583$\pm$20 km/s,
confirmed by Keel private communication) showing that past estimates
($\Delta$V$\sim$ 600 km/s) were too high. Hence the most recent encounter 
was slow unless there is a  very large transverse velocity component. 

\subsubsection{The Older Tail}

We interpret the southern tidal tail  as the older of the two because
it is broader ($\sim$9 vs. 5.5 kpc) and also shows lower optical
surface brightness (our V band estimate $\mu$=26.0$\pm$0.5 mag.
arcsec$^{-2}$) vs. $\sim$24.4 mag. arcsec$^{-2}$ (Schombert et al.
1990).  The older tail appears to emanate from the SE end of discordant
redshift NGC7320 (its full length is $\sim$100kpc if it really extends
from the SE end of NGC7320 to the OI:  Arp 1973; Arp \& Lorre
1976). This was one of the arguments used in the past for associating
the discordant galaxy with SQ (Arp 1973).  It probably passes behind
NGC7320 and connects to the region of NGC7318ab unless it really is
associated with NGC7320. N7318b would not yet have arrived in SQ when
this tail was formed.  The new HI data support this hypothesis.  HI
contours shown in Figure 5a indicate a gaseous counterpart to the old
stellar tail that indeed passes behind NGC7320 and terminates at the N
edge of NGC7320 (S edge of the shock zone).  This HI tail skirts the S
edge of the larger eastern HI cloud and appears to be morphologically
and kinematically distinct from it with a continuous velocity gradient
$\sim$ 6560-6730 km/s from east to west (S end of old stellar tail to S
edge of the shock). The large eastern HI cloud shows less ordered
velocities between 6540-6670 km/s (Williams et al. 2001).

The HI tail is connected, via the shock zone, with another (albeit
denser and more compact) HI cloud to the north (Figure 5a) with similar
velocity range (6700-6750 km/s). The connectivity in velocity is
clear.  The physical connectivity is also clear because the currently
shocked gas was most likely HI before the arrival of the NI.
We suggest that all of this material (north cloud, shock region and HI
tail behind NGC7320) represents gas that was stripped in an earlier
passage of the OI. The origin of this stripped gas might be ascribed
to an encounter between the OI and one of the elliptical
members (NGC7317 or 7318a). Certainly either of these galaxies could
reasonably have been near the north end of this tidal structure $\sim$1
Gyr ago.  The evidence is circumstantial: a) an old stellar/HI tidal
feature that most likely originated in an interaction between the OI
and one or more SQ members and b) an excess of early type galaxies in
SQ which is found in a noncluster environment where $\geq$75$\%$
late-type galaxies are expected. Indeed most nearby members of the same
large scale structure component to which SQ belongs are spirals (see
e.g.  Shostak et al. 1984). NGC7319 is the only one of the three
``core'' SQ members (NGC 7317, 18a and 19; with $\sigma_V$$\sim$0 km/s)
that can be classified as a spiral. The continuity between the northern
HI cloud, the shock zone, the HI/old stellar tail behind the interloper
motivates us to favor an origin for this structure in an earlier
passage involving the OI.  An interesting piece of evidence in favor
of a two intrusion hypothesis for production of the twin tails -- via
interaction with the OI -- involves the observation that the older
tail is not continuous (see Plate 6 in Arp \& Lorre 1976).  The fainter
tail shows a clear bifurcation or ``scalloping'' ($\sim$1.2 arcmin from
the S end of NGC7320) just beyond the part that shows overlapping HI
emission.  This structure is qualitatively consistent with the
hypothesis that the OI perturbed its older tail in the process of
making the younger one.

One cannot rule out the possibility that the most recent passage of the
OI split the HI disk of NGC7319 ejecting gas both E and W.
This possibility requires exploration with models. In this view at
least some of the gas involved with the current shock originated in
NGC7319. We do not favor this view because it would leave the older
tidal tail unexplained unless a single encounter can produce twin and
parallel tails. It would also be difficult to account for the 6700 km/s
HI cloud located NW of NGC7319. The continuity between this cloud, the
shock zone, the HI tail behind the interloper NGC7320 and the faint
optical tail motivates us to favor an origin for this structure in an
earlier passage involving the OI. This interpretation points toward
an even more significant role for secular evolution in compact groups.

\subsubsection{The Younger Tail}

The younger (narrower and higher surface brightness) tail ($\sim$40 kpc
in length) is primarily a stellar feature with HI overlapping only the
outermost 10-15kpc (Figure 5a).  The HI does not conform to the shape
of the stellar tail in the overlap region although  the tail appears
to curve along the inner side of the HI distribution where the HI
intensity gradient is steepest.  The younger tail connects directly
with NGC7319 leading us to assume that both the tail and HI cloud
originated in the ISM of this galaxy and were stripped during the most
recent encounter.  The B-band CFHT image (Figure 9a) shows that the
tail is distinct from the SW spiral arm, or broken ring, of NGC7319 and
suggests that the OI passed  from NW to SE, parallel to and just west
of, the bar in NGC7319.  Figure 9a shows at least three distinct
``streams'' of stars extending from NGC7319; one coming from the W arm,
one from the direction of the central bar and a central one (which most
likely traces the path of the OI) from the interarm region.  The
streams converge into a single narrow tail about 10-20 kpc SSE of
NGC7319.

Star-gas decoupling in tidal tails has previously been attributed to
interactions  between the gaseous component and starburst winds
(Heckmann et al. 1990) or with collisionally heated halo gas (Barnes \&
Hernquist 1996) neither of which is supported by our observations.  The
recent models show that such offsets arise naturally in low inclination
prograde encounters (Mihos 2001; see also Hibbard et al. 2000). As
observed, and expected from such models, the stellar and gaseous
components are linked in the most distant parts. The decoupling sets in
as material deeper in the galaxy is stripped with increasing effects of
dissipation on the gaseous component (Mihos 2001). This suggests that
the length of the pure stellar part of the tail should be a direct
measure of the details of the encounter (relative velocity of intruder
and orientation relative to the target disk). The data favor a low
velocity passage by the OI and at low inclination relative to the plane
of NGC7319 (see Gordon et al.  2001 for an opposite example). The data
suggest that the earlier intrusion of the OI was of a different kind
that did not lead to a star-gas decoupling.

While disturbed the spiral structure in NGC7319 is not shattered. This
observation and the fact that the stellar streams feeding into the tail
appear to pass in front of the arm, suggest that the OI made a
slow passage across (and above) the disk. The HI cloud coincident with
the end of the younger tail is the most massive detected in SQ
(M$\approx$ 4.0$\times$10$^9$M$_\odot$). It shows velocities in the
range 6540-6670 km/s (see intensity contours in Figure 5a) consistent
with an origin in NGC 7319. The cloud is better described as a ``wall''
of HI rather than a tail because it is uniformly displaced about one
galaxy diameter E and S from NGC7319.  It extends more than 50 kpc
north from the optical tail and the minimum separation between the W
edge of the cloud and the E side of NGC7319 is $\sim$30 kpc.

Little line emitting gas remains inside NGC7319 except for that
associated with the AGN. Examination of Figure 4a reveal a few weak
H$\alpha$ emitting regions associated with the stellar tail and HI
cloud:  1) two near the tip of the tail (see also Figures 10ab) and 2) a
few small regions near the north end of the HI cloud (for the brightest
see Figure 10c). These emission regions are located just inside the SQ
facing edge of the HI cloud where the intensity gradient is steepest.
One of the features in the tail (Figure 10a) was detected by ISOCAM (Xu
et al. 1999 source B). It is identified as star cluster candidate 7 in
Gallagher et al. (2001). Our Fabry-Perot measures yield a velocity
V=6617$\pm$20 km/s for this region, consistent with the HI velocities
in the area, and with an origin in NGC7319. This source shows MIR and
H$\alpha$ fluxes about one tenth as strong as in starburst A (Xu et
al.  1999) that lies at the north edge of the shock.  This implies a
star formation rate of less than 0.1 M$_\odot$/year for starburst B.
This emission region may  be heavily obscured as it is bisected by an
opaque dust lane (see Figures 4a and 10a).  The associated strong dust
feature no doubt accounts for the ISO detection of starburst B. The
emission regions in the tail are either remnants from a past star
formation event associated with the creation of the tail or new
condensations.  WFPC2 images show that the weaker emission region
(H$\alpha$ flux S$_{alpha}$= 0.4$\times$10$^{-15}$ ergs cm$^2$ s$^{-1}$
or about 0.3$\times$ starburst B) near the extreme tip of the tail
(Figure 10b) is involved with a bright segmented condensation of blue
stars that extends for 8 arcsec=3.6 kpc along the tail.  We measure
B-V=0.2 (from the WFPC2 data) for this condensation. B-V colors in the
paper were estimated relative to those reported for sources in
Gallagher et al. (2001). Such a very blue  B-V measure is found at the
extreme end of the color sequence for peculiar and interacting galaxies
(e.g. Larson \& Tinsley 1978; Schombert et al.  1990) and is certainly
consistent with the hypothesis that a very recent star formation event
occurred there.  Most of this star formation would have begun after the
tail was formed (see age estimates below). The brighter H$\alpha$
emitting region (starburst B) $\sim$15 kpc from the tip of the tail is
much more gas/dust rich and is apparently in an earlier stage of
starburst activity.

If these two regions are condensing in the tail then they are the best
candidate for tidal dwarf formation in SQ.  SQ has been frequently
mentioned as a potential site for the formation of tidal dwarf galaxies
(Hunsberger et al. 1996; Plana et al. 2000; Iglesias-Paramo \& Vilchez
2001; Gallagher et al.  2001; Braine et al. 2001). Models have
successfully produced mass concentration in excess of 10$^8$M$_\odot$
either inside, with subsequent ejection  from interacting galaxies
(Elmegreen et al. 1993) or tidally unbound debris (Gerola et al.
1983).  Such a process takes time to develop (e.g. 10$^8$ years for
significant star formation to occur; Elmegreen et al. 1993) which rules
out all emission components west of the shock where all timescales are
less than this value. Unfortunately the Mendez de Oliveira et al.
(2001) velocity measures did not cover the regions of the tidal tails.
We suggest that the local velocity gradients that they identify in or
west of the shock are largely consistent with a larger scale velocity
gradient (Section 4.2.2) that reflects the residual angular momentum in
the NI disk. Some of the HII regions south of the NI are quite large 
(D$\sim$400-500pc) but, if our interpretation
is correct, they are large because they are in the process of shock
ablation. The best, and probably only viable candidates in SQ involve
starburst B and the blue concentration located, respectively, near or
at the tip of the new tidal tail. Detection of significant CO emission
(Braine et al. 2001) provides further support for the candidacy of
starburst B near the end of the younger tail (Figure 4a and 10a). All of
the emission features on the E side of SQ fall on or near the inner
side of the HI cloud where a relatively steep velocity gradient is
observed (Figure 5a). The HST WFPC2 images suggest that all of the other 
candidate tidal dwarfs/emission regions in the tail are stars or 
background galaxies.

The most intriguing condensations are found near the north end of the
HI cloud that lies east of NGC7319 (Fig. 5a and 10c). The two brightest
objects are identified as star cluster candidates 30-31 (Gallagher et
al. 2001) but their light is dominated by line emission (e.g. they are
not detected in the H$\alpha$ IF image tuned to the NI
velocity range. They are located more than 30kpc E from that galaxy and
N of the stellar tail. The very deepest photographic images (Arp 1973)
reveal small condensations of luminous debris here as well as more
filamentary structure. Such features are so faint that it is tempting
to  associate them with galactic cirrus (e.g. Sandage 1976). This is
unlikely to be the case for the emission clumps detected in Figure 5a.
Our Fabry-Perot velocity measure for the brightest of these
condensations (V= 6627 km/s) agrees very well  with HI velocities in
the same area.  This emission region shows a curious double structure
(Figure 10c) and is superimposed on a faint linear feature (Plate 6; Arp
\& Lorre 1976) that extends towards the NE. The most straightforward
interpretation is that we are seeing the formation of extragalactic HII
regions in the stripped ISM of NGC7319.

\subsubsection{Estimating Ages for the Tails}

An important chance to advance our understanding of the OI
intrusions involves estimates of timescales.  There are several ways to
do this: 1) estimating the orbital period of the OI by using the
tails as an orbital tracer, 2)  estimating a diffusion timescale for
the old tidal tail relative to the new one and 3)  using the  colors of
the blue tails to infer a timescale by assuming that the tails  began
with a starburst event. An even more simple-minded approach involves
asking how long it would take material at the edge of the disk in
NGC7319 to reach the tip of the tail traveling at V$\sim$200 km/s
(assumed orbital velocity when inside NGC7319) which yields 
t=2$\times$10$^8$ years.

One can use the shape of the tails to extrapolate the projected orbit
of the OI on the sky.  The simplest approach for orbit
estimation involves assuming an approximately circular orbit with
projected diameter equal to the  current distance between NGC7319 and
the OI (3 arcmin$\sim$78kpc) which yields an estimated time
t=2.5$\times$10$^8$  years since the encounter with NGC7319 and about
7.5$\times$10$^8$ years since the old tail was produced (following a
possible encounter with  NGC7318a). This estimate assumed $\Delta$V=
300 km/s between the OI and SQ. The bulk of this motion is
transverse since our Fabry-Perot measure (also Keel, private
communication) indicate that there is no significant line of sight
velocity difference. The above timescale estimate represents a
reasonable lower limit. The shape of the tails, and the clear evidence
for a last perigalactic passage inside the W arm of NGC7319, raise an
interesting question about the baryonic center of mass in SQ.  The
obvious guess would be that the center of mass lies near the center of
the light distribution which would place it on or near to NGC7318a and
at least 20-40 kpc W of NGC7319. The tidal tails then place it at least
one galaxy diameter too far to the east. This may imply a very flat
mass distribution and a center of mass ill-defined or defined more by
non-baryonic than visible matter.

The old tail is fainter and more diffuse with a surface brightness
similar to the feature found in the Centaurus cluster (Calcaneo-Roldan
et al. 2000) but the conditions in SQ mimic a cluster environment in
many ways.  We assume here that the old tail was as broad as the young
one one orbit ago. It is difficult to find quantitative estimates for
the diffusion timescales of  tidal tails. Recent models suggest that
tidal tails are more easily generated in shallow potentials (Dubinski
et al.  1999). This is perhaps consistent with the previous discussion
about the implied orbit of the OI. Surprisingly, in view of
the nature of the encounters in SQ, (and compact groups in general)
close encounters are found to be more effective in generating straight
tails (Dubinski et al. 1996). The models predict that most of the
motion lies along the tail.  The streaming seen in the young tail on
CFHT and WFPC2 images certainly supports that expectation. Transverse
motions in tails are generally expected to be quite small. If we assume
a transverse velocity of 15 km/s, and ask how long it would take for
the new tail to assume the width of the older one ($\sim$4.5 to $\sim$
9 kpc), we obtain t= 2-5$\times$10$^8$ years.

It has been demonstrated that tidal tails are usually bluer than the
galaxies involved in the interactions that produced them (similar to
the colors of spiral galaxy disks from which most originate: Schombert
et al.  1990).  The bluest tails show B-V=0.2-0.4. We adopt 0.4 as a
typical value for a young tail that still shows significant star
formation. This assumption is supported by the B-V=0.4 measure
(Schombert et al. 1990) for the source A starburst (Xu et al. 1999)
involved with the ongoing collision in SQ.  The same observations
report B-V=0.57 for the younger tail consistent with a B-V=0.2 change in
color. Estimates for the timescale for such a B-V=0.2 change (Wallin
1990; Bruzual \& Charlot 1993; Calcaneo-Roldan et al. 2000) yield t=
2-4$\times$10$^8$ years. This assumes that the star formation in the
tail was quenched when the tails were formed. As discussed earlier,
parts of the tail are much bluer presumably due to star formation
initiated well after the tail was formed.  This makes the inferred
color change a lower limit to the age of the tail.  We note that
twisting is also predicted in some of the Wallin (1990) models. The
WFPC2 images show clear evidence for twisting in the younger tail with
a prominent twist being visible in the region shown in vignette A of
Figure 10.  A population of bright (M$_V$= -9 to -11) star cluster
candidates have been identified (on the WFPC2 images) in the tail with
ages (1$\times$10$^7$ to 5$\times$10$^8$ years) consistent with coeval
formation with the tail or in succeeding star formation activity
(Gallagher et al. 2001).

All estimates are consistent with an age for the new tail between 2 and
4$\times$10$^8$ years, the event that stripped the ISM of NGC 7319. If
we assume that this represents half an orbital period then the old tail
was produced 6-12$\times$10$^8$ years ($\approx$1 Gyr) ago.  SQ is
particularly interesting because it contains both old and new tidal
tails.  This allows us to make cautious inferences about the probable
evolution of a tail over a period of $\sim$1Gyr. Other tails studied so
far include those with estimated ages similar to both the old (Hughes
et al. 1991; Mirabel et al. 1991)and new (Mirabel et al.  1992) SQ
tails. The observed difference in surface brightness between the two SQ
tails suggest that $\leq$ 1 Gyr will be sufficient for the older SQ
tail to diffuse beyond detectability and, presumable, contribute to the
diffuse stellar halo which in SQ is estimated to have an integrated
luminosity $\sim$1L$_*$ (Moles et al. 1998). SQ shows similarities with
other tails: 1) a dust component, 2) the site of significant star
formation well after the tail was formed, and 3) blue condensations
near the tip of the tail ( Hughes et al. 1991; Mirabel et al. 1991,
1992). The dust feature in the SQ tail is much more well defined than
the one seen in Leo triplet. WFPC2 images show it as a well defined
feature similar to the lanes observed in the disks of spiral galaxies.
The dust feature follows the edge of the presumably associated HI cloud
and suggests that  all components of the disk in NGC7319 were pulled
(unravelled) into the tail but shortly inside the innermost part of the
dust lane the observed HI decoupling set in.  The dust lane is either
involved with or partially obscures starburst B which is a reasonably
strong MIR source (Figure 8a). Perhaps the assumed older (Rots 1978)
Leo triplet plume  indicates that this younger dust feature will
diffuse out over the next Gyr. The blue condensation at the tip of the
SQ tail is the bluests of any studied in detail so far. The weakness of
associated H$\alpha$ emission suggests that a burst of star formation
recently ended here. The WFPC2 B-band image suggests possible
decoupling of this feature from the rest of the tail while the V and I
band images show much more continuity.

\subsubsection{NGC7319 and 7320c--Spirals in Transition}

The observations point toward a significant role for secular evolution
of galaxy morphology in compact groups  with spiral members likely
being transformed into E/S0's. There is a clear analogy with the
``galaxy harassment`` process in clusters (e.g.  Moore et al.  1999).
The analogy is, of course even more appropriate for the NI
with its cluster-like velocity relative to SQ. While we cannot prove
that either NGC7317 or 7318a were originally spirals, NGC 7318b and
7320c are clearly in morphological transition.

The fact that we can still classify NGC7319 as a spiral (with a broken
ring structure) galaxy is further support for our inferences that: 1)
it has not been involved in a direct collision and 2) it was stripped
0.2-0.6 Gyr ago. The latter inference is motivated by the fact that the
spiral structure, while lacking emission regions, is still well defined
indicating that the last locus of star formation is still
recognizeable.  The same can be said about the OI
which shows residual spiral structure and is classified RSXS0 in the
RC3. Radial profiles of this galaxy reveal a bright central bulge
surrounded by a disk/ring component. This structure is  surrounded by
low surface brightness emission with residual spiral arms. Both NGC
7319 and 7320c have lost all/most of their ISM and the OI may also
have lost its disk component. The HI disk of the OI is part of the
old tidal feature discussed earlier. NGC7319 has lost all detectable
(upper limit $\sim$10$^8$ M$_\odot$) HI, the vast majority of HII
regions (expected in a typical $\sim$SBb spiral) and an uncertain
fraction of stellar mass.  A few CO clouds have been detected across
the face of NGC7319 (Yun et al. 1997, Gao \& Xu 2000; Smith \& Struck
2001). They show little correspondance with the spiral arms and may
simply involve debris above the plane of the disk and unrelated to the
galaxy.  Evidence supporting this suggestion involves the
correspondance between the CO clouds and dust patches which can be seen
in silhouette against the disk on deep images.

At this epoch NGC7319 is without an ISM that could sustain the star
formation necessary to propagate and define a population I spiral
pattern. The brightest condensation of young blue stars (B-V=0.4-0.5)
are found on the NE edge of the disk: 1) just outside the largest
H$\alpha$ + molecular gas concentration found along the line of sight
to NGC7319 and 2) on the side opposite from the inferred path of the
OI.  This condensation of H$\alpha$ and CO emission in NGC7319
(Figure 10d) is $\sim$8kpc NNE of the nucleus. CO and H$\alpha$
observations give consistent velocities V= 6800$\pm$20 km/s, with an
inferred H2 mass of 4-7$\times$10$^9$ M$_\odot$ (Yun et al. 1997; Gao
\& Xu 2000; Smith \& Struck 2001). Star cluster candidates identified
in this area (Gallagher et al. 2001) show colors consistent with ages
in the range 8$\times$10$^6$ to 1$\times$10$^8$ years.

The (V-band) luminosity of the younger tail L$_{tail}$=
0.10-0.15L$_{N7319}$ which suggests that the most recent encounter
contributed approximately that fraction of the total diffuse halo mass
where L$_{halo}$$\approx$L*$\approx$L$_{N7319}$ (R band measure: Moles
et al.  1998).  The B-band fraction would be considerably higher
because: 1)  significant {\it in situ} star formation likely occurred
after the tail was formed and 2) the diffuse halo is considerably
redder than the tail (Schombert et al. 1990; Moles et al. 1998). The OI
is undetected in all HI surveys including the sensitive observations
discussed here. In addition to morphological evidence that it is/was a
spiral galaxy we can add the results of recent unpublished 2D
spectroscopic mapping at KPNO (Keel, private communication) that
reveals weak Halpha+[NII] emission showing ordered rotation. NGC7319
and 7320c will presumably slowly transform into lenticular or
elliptical (if interactions destroy the stellar disks) galaxies.  This
will enrich SQ with early type galaxies. Such an excess of early type
E/S0 galaxies in compact groups is well established (Hickson et al.
1988) and there is little evidence that this excess might, instead,
have arisen from mergers (Zepf et al. 1991).  In another $\sim$Gyr SQ
will have no spiral members unless another intruder arrives or unless
NI loses a large amount of kinetic energy.

\subsubsection{The Seyfert 2 Nucleus of NGC7319}

The central region of NGC7319 is luminous from X-ray to radio
wavelengths.  Most of this luminosity is assumed due to the Type 2 AGN
hosted there.  A radio continuum survey of compact groups (including
SQ) revealed a deficit of global radio emission from member galaxies,
but  an excess of compact nuclear emission in the spiral components
(Menon 1995).  While most of the gas will be stripped from galaxies in
compact group environments, some undetermined quantity is efficiently
channeled into the center giving rise to active nuclei (Coziol et al
2000). More recent high resolution and s/n observations of the Seyfert
2 nucleus in NGC7319 (Aoki et al. 1999) reveal a smallscale ($\sim$ 5
arcsec) triple lobe radio continuum structure.

NGC7319 contributes more than half of the MIR/FIR emission observed
from SQ (Figure 8ab) which raises interesting questions:  1) should
this flux be added to the FIR fraction assumed due to star formation or
2) does this emission arise from dust heated more or less directly by
the nonthermal source. It is clear that the emission is thermal in
origin because an extrapolation of the radio nonthermal power-law would
not have been detected by IRAS or ISO (see radio continuum and IR
fluxes in van der Hulst \& Rots 1981; Allam et al.  1996;
Verdes-Montenegro et al. 1998; Aoki et al. 1999). Without a star
formation contribution from the nuclear region of NGC7319, SQ shows a
strong (FIR) deficit rather than an excess in contrast to the
interaction induced star formation enhancement that is observed, for
example, in binary galaxies (Xu \& Sulentic 1991). But this is not
surprising given that almost all nonstellar matter in SQ is no longer
bound to individual galaxies.  Conditions in the NGC 7319 ISM ``debris
field'' north and east of the younger tail apparently do not favor
condensation and star formation (at least in the first 0.4$\pm$0.2 Gyr)
-- except for a few isolated clouds inside and north of the younger
tidal tail (section 4.1).

Many Seyfert 2 nuclei show evidence for a near nuclear star formation
enhancement. In fact some may not be AGN at all (Dultzin-Hacyan \&
Benitez 1994).  In the case of NGC 7319 there is little evidence for a
nuclear starburst. HST images of the nuclear region (the central
2$\times$2 arcsec: Malkan et al. 1998) show complex bright and dark
spiral  structure but no evidence for star formation condensations.
Radio continuum (Aoki et al. 1999) and supporting optical slit
spectroscopy (Aoki et al. 1997) show evidence of complex internal jet+
triple-lobe structure on a scale of 10 arcsec but no signature of star
formation. The high surface brightness structure seen in the inner kpc
by WFPC2 is no doubt the source of the line emission seen on slit
spectra.  But this region is dominated by forbidden emission rather
than a starburst signature.  We compared the stellar psf with the
nuclear H$alpha$+[NII] emission in NGC7319. The psf for stars in and
near NGC7319 on the average of our Calar Alto R-band continuum images
(taken immediately after the emission line images) give a consistent
FWHM=3.5 arcsec (the seeing was not good).  The H$\alpha$+[NII]
emission line psf gives a similar psf (3.5-4.0 arcsec) except for a
wing  extending about 10 arcsec to the SW. The wing corresponds to the
mostly forbidden emission studied by Aoki et al. 1997). The seeing disk
therefore matches well the field imaged by WFPC2 where no evidence for
significant starburst activity was found.

There is conflicting evidence for the presence of molecular gas in the
nucleus of NGC7319. Demonstration of significant molecular emission
from the nucleus would support the argument that significant star
formation was occuring there. Yun et al. (1999) detect a small cloud
about 2kpc S of the nucleus but Xu \& Gao (2000) place this source on
the nucleus. We favor the former interpretation because the former
(somewhat higher resolution) centering coincides with an optical dust
patch seen in silhouette and the CO emitting region is clearly extended
and not concentric with the nucleus. Thus it may well be a projected
cloud of tidal debris rather than a concentration of molecular gas in
the immediate nuclear region. In any case, the mass of molecular gas
there is $\leq$2$\times$10$^8$M$_\odot$ inconsistent with the bulk of
the observed MIR/FIR emission originating in a hidden nuclear burst.
If one interprets the Seyfert 2 nucleus of NGC7319 as a product of the
most recent encounter with the OI then: a) only a small quantity of
gas escaped the stripping event, 2) this small quantity was quickly and
efficiently  channeled into the innermost regions, 3) it gave rise to
an AGN with associated small-scale radio lobes and an optical jet
structure involving shocked gas, 4) little star formation is/was
involved with this process and 5) the dusty nuclear environment is
likely heated more or less directly by the  power-law continuum of the
AGN which reenforces our inference that the FIR signature of star formation 
in SQ is very weak. 

\subsection{SQ: The Ongoing Collision}

\subsubsection{The SQ ISM}

We assume that NI was not inside, or near, SQ when the interaction
events described in the previous section transpired. Reasonable
crossing times (t$_c$= 2--8$\times$10$^7$ years) for NI are
5-10$\times$ less than the estimated time since the last encounter
between SQ and OI.  Any shocks generated by past intrusions would have
cooled, unless they involved very low density
(n$_e$$\sim$10$^{-2}$--10$^{-3}$ cm$^{-3}$) gas, and most
interaction-induced star formation activity would have ended. We
observe an X-ray and H$\alpha$+[NII] extension  (Figures 2c and 4a)
from the N-S oriented shock zone towards NGC7319. Its orthogonal
orientation with respect to the shock motivates us to interpret it as
related to the ongoing events.  The most important inputs into our
inferences about ongoing events involve:  1) the configuration of the
stripped ISM in SQ before the arrival of NI and 2) the trajectory and
overall structure of NI.  We suggest that Figure 5a provides the most
reasonable estimate of the immediate pre-shock SQ ISM. This
superposition of HI (contours) on H$\alpha$+[NII] emission represents
all warm and cold gas (except molecular) in the SQ velocity range.
[NII] emission (Figure 2d for the brightest part of the shock and
Figure 4a for more extended diffuse emission) also traces the location
of the hot gas. If one replaces the regions showing shocked gas with HI
then one should have a reasonable idea of the pre-shock ISM. 

Velocities of stars and gas associated with SQ span a range from
6400-7000 km/s (Figures 4a, 5a and 7a; Table 2).  The gas surrounding
NGC7318ab spans the velocity range 5400-6000 km/s (Figures 4b, 5b and
7b).  Therefore NI velocities are found in, or west of, the shock while
SQ velocities are found in, or east of, the shock.  The only exception
to this rule involves a radio, X-ray and H$\alpha$+[NII] extension
(towards an emission condensation with an uncertain Fabry-Perot V$\sim$
7000 km/s) towards the extreme NW of SQ. The SQ and NI emitting regions
are reasonably well separated in the IF images shown in Figures 4ab.
Some redundancy exists due to overlap in the filter responses in the
velocity range 5900-6400 km/s (Figure 6).  Fabry Perot velocities are
given in Table 2 for all of the detected emitting regions which are
numbered on the contour plots shown in Figure 7.

We find HI clouds with unambiguously SQ velocities (Figure 5a): a)
north of the NI, b) south of the NI, passing behind NGC7320 and
superimposed on the old optical tail, as well as  c) the largest cloud
located ESE of NGC7319. Clouds b) and c) appear to be structurally and
kinematically distinct with b) part of an old tidal feature, perhaps
involved with the stripping of NGC7318a or 7317, and c) involved with
the more recently stripped ISM of NGC7319. We suggest that clouds a)
and b)  are linked by the shock region where we find H$\alpha$+[NII]
emission (Figure 5a).  Gas in the region of the shock would have been
largely cold (like HI clouds a) and b))  before the arrival of the NI.  
This is a reasonable hypothesis because of the long
estimated time since the last encounter and because of the small amount
of H$\alpha$ emission observed in optical and HI tidal features
produced in past events (making it likely that little or no H$\alpha$
in the shock region, with SQ velocities, is related to past events).
One possible problem with this ``linkage'' involves the higher density
and roughly circular shape of the northernmost HI cloud. We argue that
the velocity continuity argument carries more weight. And the fact
that, at least, a projected linkage is undeniable in the absence of the
current shock conditions. Finally, the HI in tidal features is clumpy
and the north clump would be nearest the impact point that produced
this assumed tidal feature.

The high velocity passage of the NI through the stripped gas
(the gas between components a and b above) gives rise to the shock. All
relevant timescales connected with this event are less than 10$^8$
years:  a) NI crossing time, b) decay of the radio
synchrotron emission ($\sim$8$\times$10$^7$ years; Van der Hulst \&
Rots 1981), c) the shock cooling time, d) the age of Starburst A in SQ
ISM on the north edge of the shock (1--2$\times$10$^7$ years; Xu et
al.  1999) and e) the ages of the bluest star cluster candidates
identified in SQ (consistent wiith ages 5-7 $\times$10$^6$ years).
Thus we assume that this is an ongoing collision. The implication of
this assumption involves evidence in our data that: 1) the ISM of the
NI is not yet completely stripped and 2) we are seeing
NI HII regions in the process of ablation (see Section 4.2.2).

The NI may have only grazed the HI feature which is now
partially shocked, given that its extreme line-of sight velocity
relative to SQ makes it unlikely that it has a very large transverse
component. The fact that we detect both SQ and NI emission along
the same line of sight at various places along the shock support this
interpretation. If the elongated shock arose from a transverse, rather
than line-of-sight, component in the NI velocity we might
expect to observe an eastward displacement of the SQ emission regions
relative to those with NI velocities.  Existing data finds
them along the same line-of-sight. In any case we can not view the
preexisting HI feature as a ``wall'' but instead as an old tidal
feature with well defined extent. Of course we cannot rule out the
existence of more extended HI (or X-ray emission) across a much wider
part of SQ. In fact the large line of sight velocity difference makes
it likely that the bulk of the shocked gas should be along the line of
sight if there is any gas to shock in that direction. If present it may
have a column density too low for detection (in HI or X-ray).  The
``missing'' X-ray emission (see Section 2.1)  may originate in this line
of sight component. Evidence for a more diffuse optical emission
component comes  from Figure 4a where we see  very extended and faint
emission in addition to that associated with the N-S shock zone.  It is
reasonable to assume that most of this is probably diffuse [NII]
emission.

Emission from the stripped ISM in SQ will involve two components: 1)
photoionization recombination emission (e.g.  H$\alpha$) from denser
emission regions associated with star formation and 2) forbidden
emission from low density shocked gas (e.g.
[NII]$\lambda$6583$\rm{\AA}$). Post-shock emission regions can be
distinguished from pre-shock NI emission by the considerable difference
in velocity. Shocked gas will brake from near 5600-6000 km/s to
6400-6700 km/s. Optical emission from the shocked gas should follow the
radio continuum and X-ray emission and this is confirmed in Figures
2abcd. Our Fabry-Perot measures along the  shock are often difficult to
interpret. The forbidden [NII] emission is often so broad that it
cannot be isolated as a line even with 950 km/s free spectral range.
Pre- and post-shock emission regions are both detected in several
places along the shock zone where the Fabry Perot measures show two
emission line peaks, one at NI (5400-6000 km/s) and the other at SQ
(6300-6400 km/s) velocities.  We may be detecting recoil in the shock
zone because SQ velocities are $\sim$300 km/s lower than unshocked HI N
and S of the shock.

Comparison of ISO MIR 11.4 (Figure 8a) and 15$\mu$m (Xu et al. 1999 and
contours in Figure 8b) with the H$\alpha$ images shows a weak
correlation in the sense that only weak evidence for a MIR ridge or
enhancement is seen along the shock zone.  This is in marked contrast
to the situation for galaxy pairs where FIR and especially MIR
correlate strongly with H$\alpha$ emission (Xu \& Sulentic 1991; Toledo
et al. 2001, Domingue 2001).  MIR emission is weaker in the shock
region than expected, for example from the shock/starburst A ratio for
H$\alpha$ emission. The ratio of continuum subtracted H$\alpha$ flux
(a 100 arcsec$^2$ region located 40 arcsec south of starburst A for the
shock measure) is R$\sim$0.3-0.4$\pm$0.1 while the corresponding
11.4$\mu$m ratio is R$\sim$ 0.14$\pm$0.1.  Therefore we suggest that
MIR emission may be suppressed in the shock region. If this is true
then MIR/FIR measures will underestimate the star formation rate in the
post shock gas. The weakness of the MIR emission along the shock may be
telling us that much of the dust has been diffused and/or destroyed by
sputtering in the shock region

One very bright emission region is seen at the immediate edge of the
N-S shock. ISOCAM starburst A with V=6670 km/s is a strong source of
H$\alpha$+[NII], MIR (Xu et al. 1999), HI and CO (Smith \& Struck 2001)
emission.  This is a complex region where both HI and CO spectra show
two strong and narrow velocity peaks (6000 and 6700 km/s).  Our
Fabry-Perot measures indicate that all of the emission regions north of
starburst A originate in the NI and account for the 6000 km/s
detections. Starburst A is either: a)  a condensation of gas at the
very edge of the most intense shock region (the SQ ISM at starburst A
has been compressed but not shocked) or b) it is an unusually
dense region that has cooled out of the shock more quickly than
adjacent regions. Local conditions have favored the conversion of much
of the gas into molecules M$_{HI}$= 1.5-5.0$\times$10$^9$$M_{\odot}$
and $M_{H2}$=1$\times$10$^9$$M_{\odot}$)(Smith \& Struck 2001). It is
difficult to give a reliable estimate for the H$\alpha$ intensity or
equivalent width from our calibrated interference filter images. The
relative strengths of  starburst A and B (Xu et al. 1999) are more
different than the EW values suggest (e.g. the starburst B/A EW
ratio=0.74 while the flux ratio= 0.11).  The apparent EW similarity
arises more because of the relative intensity of the adopted
normalizing continuum.  Both intensity and EW values also depend upon
the correction for [NII] contamination which is maximally uncertain in
the SQ shock region. Starburst A appears to be heavily reddened based
on Figure 8b where it is the only bright emission region that is
distinctly red in color (B-R=1.3 compared to B-R=0.5-0.7 in emission
regions just north of it; Mendes d'Oliveira et al. 2001). In fact this
appears to be the effect of strong H$\alpha$+[NII] emission in the R
band because we derive B-V=0.5$\pm$0.05 from the WFPC data. In the
absence of high s/n slit spectra, the MIR data (Xu et al. 1999) are a
more reliable estimator of the star formation rate in this case (B/A
15$\mu$m flux ratio= 0.15 much more similar to the H$\alpha$ flux
ratio). Starburst A at the edge of the shock region is almost
10$\times$ more intense than starburst B at the end of a tidal tail
produced $\sim$4$\times$10$^8$ years ago.

\subsubsection{New Intruder ISM and Configuration}

The main goal of this section is to discover the properties of the NI.
It is difficult to construct a 3D model of the galaxy because there is
significant if not complete stellar--ISM decoupling. The stellar disk
is traveling through SQ with such a high velocity that it  probably
remains little disrupted.  Identifying the stripped disk of the NI is
particularly difficult because: a) it will be projected on the bright
galaxies and tidal debris in SQ and  2) parts of the NI may pass around
both sides of the HI clouds in its path and avoid being shocked at
all.  Our conclusion that the NI ISM is not completely shocked is well
supported by the F-P and HI data which resolve considerable confusion
about the nature of the nonstellar matter around NGC7318ab. N and SW of
NGC7318ab we see chaotic concentrations of emission regions in
H$\alpha$ (Figure 4b) as well as the CFHT B-band image (Figure 9a).
They are even more clearly seen on the CFHT B-R image because of their
blue color (white in Figure 9b) relative to the much redder underlying
stellar light.

The bluest concentration of emission regions (we measure B-V=0.3-0.5 
$\pm$0.05 on the WFPC2 images) is seen just north and likely passing
in front of starburst A (Figure 10e).   The equivalent
emission regions to the S and SW of NGC7318ab are fainter (excluding
the 4 largest and brightest condensations nearer to the center of the
NI) which may indicate that they are on the far side of the NI
and more affected by the complex dust lanes and patches that can be
seen especially well on the CFHT and WFPC2 images. The optical feature
extending northward from NGC7318a toward the blue condensations
(variously referred to as a tidal tail or arm) is confirmed as NI 
material by the detection of a few HII regions (at both ends)
with velocities that are internally consistent and that connect
smoothly to the northern emission regions. Figure 5b shows the HI
clouds with NI velocities: a) 5600-5800 km/s located SSW of the
nucleus and b) 5960-6020 km/s located N of the nucleus. Wherever HI
clouds overlap (Figure 5b) we find excellent agreement between HI and
Fabry-Perot velocities. There is evidence for a residual rotation
pattern in the NI emission regions over the range 5400-6000 km/s:
from 5400-5700 towards the S and SW, increasing to 5800-6000 km/s
towards the W and N.  Velocity gradients in the two HI clouds are
consistent with this trend. The two distinct H$\alpha$+HI
concentrations suggest that the ISM of the NI has been split, in
the sense that the gas between them has been shocked/displaced  but the
angular momentum associated with this residual ISM has not been
dissipated--further support for the hypothesis that the collision is
ongoing.  NI emission regions that collide with the SQ ISM at
such high velocity will be shocked and ablated. At the same time HI
will be rapidly heated in the shock.  We therefore conclude that the
complex HI+H$\alpha$ structure W of the shock involves HII regions
either not yet shocked or ones that have missed the shock entirely.
No emission around NGC7318a (nothing W of the shock--except possibly at
the extreme NW edge of SQ) was found to have a velocity within 600 km/s
of the nuclear velocity of NGC7318a. The lack of any emission regions
attributable to NGC7318a is consistent with its classification as an
elliptical galaxy (see also Moles et al. 1998).

The deep CFHT images offer the best chance to make a reasonable
estimate of the size and shape of the NI. On the B and R
IMAGES we can detect much faint structure that is roughly symmetric
about the central bulge of the NI.  The NW edge of the disk
is particularly well seen with much faint flocculent structure. This
structure trails in the same direction as the arm that extends from the
edge of NGC7318a towards the HII condensation north of starburst A. The
SW extent of the NI is indicated by the HI cloud and
H$\alpha$ concentration (Figure 5b). The distribution of HII regions
there is extremely chaotic. Pairs of very bright HII regions on the SW
side of the NI flank an apparent break in a ring or spiral
arm. Emission regions in the gap between the bright emission knots are
displaced towards the SW by several kpc. All of this region is
enveloped by one of the NI HI clouds.  Perhaps this region passed
directly though NGC7318a. The northernmost emission regions belonging
to the NI extend north of starburst A and eastward from the shock
near to the  north edge of NGC7319.  Our B-R image (Figure 9b) may
provide the most valuable clue towards unravelling the outermost spiral
arms in the NI.  It shows a collection of faint blue
condensations emerging from the SE side of the central bulge/bar and
passing across and south of the brightest HII regions mentioned above.
This faint chain of emission regions  passes across the N edge of
NGC7320.  This feature connects with the HII concentration on the SW
edge. One region within NGC7320 (region B1 of Arp 1973) and all regions
to the SW are confirmed to show NI velocities. We suggest that
these emission regions trace the outermost NI spiral structure.
In fact this feature connects more smoothly with the east end of the
NI bulge/bar structure than does the bright string of emission
regions closer to it. The latter appear to be most directly involved
with the shock at this time.

The ellipse superimposed on Figure 9b encloses all of the
structure that belongs to the NI on the basis of kinematic or
morphological continuity. The outermost southern spiral arm seen on the
B-R image has a reasonably symmetric counterpart on the opposite side
that involves the arm passing behind or in front of NGC7318a and
extending north of starburst A. CFHT and HST images suggest that this
material is residual spiral structure--filamentary structure involving
both bright and dark lanes are clearly visible here. All observations
indicate that the NI is/was a large spiral galaxy especially
given: a) the amount of atomic and molecular gas that can still be
assigned to it and b) given the residual rotation found in the velocity
measures ($\Delta$V$\sim$600 km/s). If we assume that the arms trail,
the sense of rotation sees the N side nearest and receding. The
major axis diameter of the ellipse in Figure 9b is $\approx$2.6
arcmin$\sim$65kpc and the minor axis $\approx$1.6 arcmin$\sim$ 40 kpc
which implies an inclination to the line of sight in the range of 
30-40$^{\circ}$.

The fact that: 1)  SQ shows an HI deficit despite the number of well
defined HI clouds found there (Figures 5ab; Williams et al. 2001) and
2) that almost 1/3 of our X-ray photons cannot be assigned to a
discrete source provide some support for the hypothesis that there may
be shocked gas across the broad extent of the NI (our upcoming
NEWTON observations will be more sensitive to such emission). The well
defined arc that defines the shock at X-ray, radio continuum and
optical emission argues that part of the NI, the stripped part,
maybe E of the shock. Alternately, if the bulk of its motion is along
the line of sight, one can argue that the edge of the NI is
brushing past an old tidal HI feature as discussed earlier and
suggested by Figure 5a.  In this scenario a significant part of the
NI may never be shocked. The strongest argument for a significant
component of motion towards the NE involves the extensive structural
smearing seen in luminous and dust-related absorption  features on the
N and E side of the NI (Figure 10f).  One can see long luminous
filaments extending for more   than 20-30 kpc in Figure 9a. They extend
from near the center of the NI towards the northeast.  A
prominent dust lane is the southernmost of these features which extends
from the NI directly towards the nucleus of NGC7319 (Figure
9a; also visible on the photographs of Arp 1973).  This structure does
not look like residual spiral arms because it is too filamentary. The
closest analogy one can find involves the spokes of the Cartwheel
galaxy (Struck-Marcel \& Higdon 1993; Struck et al. 1996). Perhaps the
analogy to the Cartwheel is reasonable if the disk of the NI
has passed directly through NGC7318a.

We earlier proposed that the intrusion of the NI was not only recent
but ongoing. Aside from the timescales cited earlier the bright
emission knots south of the NI and NGC7318a may provide the
most direct evidence.  These features complicate any interpretation of
the overall pattern of the NI. They appear anomalously bright
and large (the largest show diameters of 400-500pc) compared to other
emission regions with NI velocities. One must concede to Arp
(1973) that there are HII regions in SQ both larger and more luminous
than any in NGC7320 which is almost ten times less distant. The SE
string of these emission regions appear to connect to the E end of the
bar in the NI. Our CFHT B-R image suggest that instead a fainter
arm passes from the bar and defines the south edge of the NI
disk. The very bright features may perhaps be part of a disrupted
internal ring.  Whatever their origin we suggest that they represent
NI emission regions in the process of ablation at the shock
boundary.  At this boundary HII regions will expand in directions
perpendicular to the direction of motion assuming a pancake like shape
as they become optically thin and dissipate. Little theoretical
modeling exists for such a scenario possibly because such a situation
is likely to be rare.  The more common but equivalent scenario
involves emission knots in a hot stellar wind. In either case a large
amount of NI kinetic energy will be converted into mechanical
energy at the shock front. This energy will heat the SQ ISM as the
NI emission regions are ablated giving rise to the observed X-ray
and forbidden optical emission. At the same time compression of the ISM
and associated magnetic fields will increase the density of nonthermal
electrons giving rise to nonthermal radio emission. Some of the
elongated emission features in the shock front resemble the expected
shape of Rayleigh-Taylor instabilities or ``fingers'' that are expected
in such a medium if the magnetic field lines are more or less
perpendicular to the shock front and are compressing the gas against
motion along the shock.  A candidate instability feature is shown near
the top of Figure 10h but many can be seen on archival WFPC2 images.

If our interpretation is correct then we have direct evidence for a
shock component along the line of sight because the largest HII regions
are west of the shock. If they are ablating then we know that the shock
geometry is much more complicated than implied by the N-S ``arc''.
Surprisingly little high s/n spectroscopy exists for emission regions
in the NI.  Two slit spectra, one published (Gallagher et al.
2001) and one unpublished (J. Gallagher, private communication) exist
for the region of the bright knots between the NI and
foreground NGC7320.  These spectra show typical narrow emission line
emitting HII regions with NI velocities. They also show very
broad and diffuse emission regions with SQ velocities. Some evidence is
also seen for more normal HII regions with SQ velocities.  We interpret
these as pre-shock, shock and post-shock features. The published 10m
HET spectrum (Gallagher et al. 2001) intersects Table 2 emission
regions 1, 2 and 3 and then passes just SE of knot 6. The former three
regions show somewhat broad emission lines with unambiguously NI
velocities. More diffuse gas between knots 3 and 6  shows velocity
smeared emission (possibly at V $\sim$6050 km/s  intermediate between
the NI and SQ) presumably at the shock front.  The region just SE
of knot 6 (which has an NI velocity), shows weak somewhat diffuse
emission at an SQ velocity. Our IF images (Figures 4ab) show strong SQ
and NI emission in this region. This is a part of the shock front
where ongoing ablation is occuring. North of this region and inside the
emission arc we see possible evidence of HII regions already destroyed
in the form of diffuse disk- or ring-like features (see Figure 10h and
wider field archival WFPC2 images). The region where the HII region
``ghosts'' are seen lies directly east from the central regions of the
NI where an X-ray, radio and [NII] emission peak is observed.
An alternative interpretation might view these features as supernova
remnants.  The SQ-NI interface appears to be an ideal laboratory
for studying the ablation of emission regions in a galaxy ISM.

\section{SUMMARY AND IMPLICATIONS}

\subsection{Galaxy by Galaxy}

Proceeding from W to E we summarize our interpretation of the evolutionary
history for each galaxy in SQ.  While some of these results were reported 
previously, most or all of our inferences benefit from the first-time 
unraveling of HI and  HII velocities (from the Fabry-Perot and HI measures) 
throughout SQ.   

\begin{itemize}

\item{\bf NGC7317} is an elliptical member linked to the rest of the group 
by diffuse stellar light (Moles et al. 1998). There is no evidence for its 
involvement in any of the events discussed above.  

\item{\bf NGC7318a} is an early-type galaxy adjacent to the NI. There is a
multi component tidal tail (from N to S: HI--shock zone--HI tail+old
optical tail) located  E of this galaxy that may point to an earlier
metamorphosis from spiral to elliptical morphology. It is interesting
that NGC7318a is a stronger source at X-ray and radio continuum
wavelengths while NGC7317 and 7318b are not.  Published B, V and R
magnitudes for these three galaxies (in the case of the NI we are
referring to the luminous elliptical-like bulge component) differ by
less than 0.5 while they show almost identical colors  (Hickson et al.
1989; Schombert et al. 1990; Moles et al.  1998). WFPC2 images shows
dust lanes crossing near the nucleus.  This galaxy may be a good
candidate for infall of tidal debris since the NI disk likely passed
through it.

\item{\bf NGC7318b}, the NI, is a large gas rich spiral galaxy now
entering SQ for the first time with line of sight velocity near 1000
km/s.  It is entering the group from behind with a component of motion
towards the ENE.  Half or more of its ISM is now stripped. The residual
ISM is split into two clouds of HI, HII regions and molecular gas with
mean velocities near 5700 and 6000 km/s.

\item{\bf NGC7320} is a foreground Sd dwarf projected on the southern edge of SQ and
directly upon the northern extension of the older tidal tail. HII regions 
associated with both SQ and the NI overlap its N end.  

\item{\bf NGC7319} is an SBb spiral (with Sy2 nucleus) SQ member that has lost most
of its ISM. The new tidal tail appears to trace the passage of the OI
from NW to SE. In the last intrusion the OI passed just W of the bar
in NGC7319 and above (or below) the disk. All detectable HI from NGC7319 is now
located in a complex cloud displaced almost one galactic diameter ESE. One or
two candidate tidal dwarf galaxies may be forming in this largely quiescent
stellar and gaseous debris. Residual H$\alpha$ and (significant) H$_2$ gas
are found in (or projected on) the NE side of NGC7319 opposite the path of the
OI. Some of the gas that was not stripped may have fallen into the
nucleus therebye triggering the AGN. It is in transition from spiral to
lenticular morphology.  

\item{\bf NGC7320c}, the OI located one group diameter ENE, shows
evidence of a ring and spiral arms, but no trace of HI and weak optical
line emission. It is interpreted as the OI because both
optical tidal tails curve in its direction. It lost most of its ISM
during a passage through SQ and is now bound to the group. It is in
transition from spiral to lenticular morphology.

\item{\bf SQ} Interactions in SQ are of two kinds: 1) non-impulsive
($\sigma$$_V$$\sim$ 0, line of sight, for NGC7317, 7318a, 7319 and
7320c) quasi-continuous interactions between bound  members and 2)
episodic highly impulsive intrusions by neighbors from the associated
large scale structure. It may be unusual that two intruders have
visited SQ within the last Gyr but the frequency will depend upon the
richness of the associated large scale structure and the mass of the
attractor. In terms of galaxy surface density, SQ is not unusual
(Sulentic 1987). Nonimpulsive interactions will create halos, possibly
ignite AGN, and dissipate angular momentum while the more violent
collisions can modify member galaxies morphologically (ISM stripping,
disk disruption, accelerating the halo building process). Following
this definition, both the OI and NI have been involved in interactions
of the second kind. Either the OI lost a large amount of kinetic energy
in the last visit or the close encounter with NGC7319 was slow  and
remarkably efficient. The young tail and associated HI cloud supports
the latter hypothesis. If the ISM of galaxies in SQ was largely lost to 
impulsive encounters then it will be dangerous to extrapolate the SQ 
experience to other groups, unless the DM haloes foster rapid and 
efficient dissipation of intruder kinetic energy.

SQ has survived significantly more than 1 Gyr without any evidence for
onset of merging.  As far as it goes this suggests that mergers may
be very rare among compact groups.  They must occur before the
disruption/stripping process is too advanced or they will be unable to
achieve extreme or even unusual IR luminosities (Borne et al. 2000). SQ
suggests that the  proximity and strong interaction of 4-5 galaxies
does not always lead to rapid coallescence although it does lead to
efficient stripping of component ISMs. The latter results in depressed
rather than enhanced star formation activity. It is unclear what role
the episodic intrusions play in retarding the coallescence or whether
the fate of the group is determined mostly by the properties of the
dark matter halo thought to surround such groups. The observations
suggest that it accounts for $\sim$90\% of the mass (to accelerate high
velocity intruders) and that it is distributed very smoothly resulting
in a very diffuse, rather than cuspy,  potential.

\end{itemize}

\subsection{Implications for High Redshift Phenomena}

Compact groups manifest the galaxy formation and evolution processes
that are invoked to describe and explain what we see at high redshift.
Interactions and the effects of interactions are believed to be more
frequent in the past (e.g. Wu \& Keel 1998). Compact groups as the site
of extreme interactions at low redshift may therefore be useful local
analogs that can be studied in greater detail. We consider the
implications of SQ to various topics often discussion in a high
redshift context. 

\begin{itemize}

\item{\bf 1) Infall processes and structure formation in the Universe:}
SQ suggests that compact groups form by the sequential acquisition of,
sometimes high velocity,  intruders from the associated large scale
structure. Other Hickson groups also show internal high velocity
(likely) intruders and potential intruders just outside the isolation
boundary (Sulentic 1987). We can identify with some confidence the two
most recent (within the past 1Gyr) arrivals in SQ.  While not cataloged
as a member of SQ the OI is  now likely  bound to the system. The NI is
likely a giant field spiral that we find in mid passage.  The NI shows
a cluster like velocity relative to SQ.  If SQ is responsible for this
high infall velocity then a significant nonbaryonic mass component is
required because the baryonic mass fraction in SQ is an order of
magnitude too low to serve as an efficient attractor (Moles et al.
1997). Limits on the rapidity of formation and frequency of occurence
of this process will be set by the number of observed groups and the
density of galaxies in their local environment.

\item {\bf 2) Dynamical evolution:} SQ shows classic signs of repeated 
interactions such as tidal tails and diffuse light. SQ suggests that 
interaction with existing and incoming members leads to the gradual
stripping of both gas and stars from group members. Two recognizeable
spirals in SQ are almost totally stripped:  NGC7319 and the OI are
almost certainly evolving into E/S0 systems. SQ suggests that compact
groups evolve from predominantly spiral to early-type systems.
Early-type rich compact groups are found locally but their numbers, as
evidenced by the Shakhbazian groups (e.g. del Olmo et al. 1995), appear
to have been larger at higher redshift. The early-type rich compact
groups would be the highly evolved analogs of SQ that are most
resistent to merging.  HCG79 (Seyfert's Sextet) may be the best low
redshift example. If such groups indeed originated from gas rich spirals 
then the most intriguing question is  ``where did the gas go?''.

\item {\bf 3) Star Formation and ULIRGS:} Compact groups do not show
high average levels of star formation as inferred from their MIR/FIR
emission (Sulentic \& deMello Rabaca 1993; Verdes-Montenegro et al.
1998). This is in contrast to what is seen in pair samples found in
similar environments (e.g.  Xu \& Sulentic 1991). Dynamical evolution
in SQ-like groups will leave most gas either too hot or too cold to
support much star formation. In SQ the strongest residual starburst
activity is of an unusual kind --modest starbursts in tidal debris.  No
Hickson group shows extreme IR properties (LIRG or ULIRG). Reasonably
strong IR emitting groups are often dominated by an AGN. In the case of
SQ there is no evidence that even that IR emission is driven by star
formation. The rare examples of high redshift ULIRG compact groups
(Borne et al. 1999) are likely different from the average Hickson (i.e.
nearby) group. They are interpreted as multiple mergers {\it in
flagrante delicto} while SQ  shows no evidence for merger activity and
much evidence for systematic metamorphosis from a spiral to an
early-type rich system. Unless an ULIRG can form from a group that is
largely stripped, they must be the most ``unlucky'' compact groups
where component galaxies arrive at about the same time (avoiding
systematic disruption of the individual DM haloes) and quickly evolve
to coallescence. Such groups are rare.

\item {\bf 3) Feedback and the formation of AGN:}  SQ contains a
stripped spiral with Seyfert 2 nucleus that may have arisen because of
the violent dynamical processes occuring there. The quasi-continuous
nature of the tidal torques in compact groups may more efficiently
(than in pairs) channel gas into the nuclear regions of group members.
This process, or feedback of stripped gas, could give rise to the
excess of nucler radio sources (Menon 1995) and type 2 AGN (Coziol et
al.  2000). It is perhaps too soon to expect more systematic feedback
of stripped gas into the component galaxies of SQ.  NGC7318a may be 
the best candidate for tidal feedback since it is in the path of the NI disk.
The general HI deficit found for the groups (e.g.
Verdes-Montrenegro et al. 2001) may be evidence that such feedback is
very slow and inefficient.

\item {\bf 4) Merger phenomena:}  A conservative age between 1-2Gyr can
be assigned to the identifiable episodes of intruder activity in SQ.
SQ is dynamically evolved and all members except the ongoing intruder
have lost their ISM.  Unless significant gas feedback occurs it is
therefore difficult to envision a fate for SQ as an infrared bright
(ULIRG) merger (Borne et al 1999). If one wants to make an ULIRG then
one must do it quickly before massive stripping occurs but after the
dark matter halo is disrupted. If SQ is typical of compact groups then
this requirement is rarely fulfilled.  Yet the velocity dispersion in
SQ $\sigma_V$$\sim$0 (excluding the unbound NI) is fairly common in
local groups. Evolving ideas about massive and diffuse DM halos around
the groups may also be converging with observations that suggest that
even low velocity dispersion groups can resist merging indefinitely
(Athanassoula et al. 1997).
\end{itemize}

\section{Acknowledgements}

One of us (JWS) acknowledges hospitality and support from IAA, OAB and
IdA-UNAM during the course of this work and support from NASA-JPL
contract 961557. We acknowledge helpful discussions and assistance
from  I.  Fuentes-Carrera, A. Iovino, A. del Olmo and J. Perea. R.
Tuffs is acknowledged for assistance with ISO data reduction.  We thank
C.  Mendes de Oliveira for providing her CFHT images. The HST WFPC2
data was obtained from the multimission archive at the Space Telescope
Science Institute (MAST). STScI is operated by the Association of
Universities for Research in Astronomy Inc. under NASA contract
NAS5-26555.  LVM, is partially supported by DGI (Spain) Grant
AYA2000-1564 and Junta de Andaluc\'{\i}a (Spain). MR acknowledges
grants 400354-5-2398PE of CONACYT and IN104696 of DGAPA-UNAM, MEXICO.
D.D-H. acknowledges support from grant PAPIIT IN115599, UNAM
The ROSAT project is supported by the German Bundesministerium f\"ur
Bildung, Wissenschaft, Forschung und Technologie (BMBF/DLR) and by the
Max-Planck-Gesellschaft (MPG).

\vfill\eject

\begin{figure}
%\plotone{sulentic.fig1.new.ps}
\caption{The six panels provide finding charts for each galaxy in SQ
where the recession velocities and aliases are also noted.
Other important features discussed in the text are identified. New and old tail in the 
lower right panel are, respectively, synonymous with younger and older tail in the text. 
\label{fig1}}
\end{figure}
\vfill\eject

\begin{figure}
%\plotone{sulentic.fig2.new.ps}
\caption{Comparison of the X-ray maps from the new HRI observation of
Stephan's Quintet with radio continuum and [NII] line + red continuum
emission: UPPER LEFT (a): an adaptively smoothed HRI image with superposed 
contours. UPPER RIGHT (b): the adaptively smoothed HRI contours superposed on 
a 21cm radio continuum image (18 arcsec synthesized beam) with comparable 
effective resolution.  LOWER LEFT (c): a Gaussian smoothed ($\sigma=4''$) 
HRI image. LOWER RIGHT (d): an interference filter image centered at $\lambda$
6731$ \rm {\AA}$ (FWHM=10$ \rm {\AA}$) sensitive to [NII]$\lambda$6583$\rm{\AA}$
(+ continuum) emission in the SQ velocity range 6460-7000 km/s. 
\label{xmaps}}
\end{figure}
\vfill\eject

\begin{figure}
%\plottwo{sulentic.fig3a.eps}{sulentic.fig3b.eps}
\caption{LEFT: Plot of the regions used in Table 1 to derive fluxes of 
the different components and of the residual emission.  RIGHT:  Radial 
distribution of the total emission from SQ, azimuthally averaged in 
two arcs oriented NS and EW, in concentric annuli about the X-ray 
peak roughly in the middle of the elongated N-S extended source.  
The adopted background level is also shown. The excess at r$\sim 1'$ 
is due to the Seyfert nucleus.  Data are from the new observation only.
\label{xprofile}}
\end{figure}
\vfill\eject

\begin{figure}
%\plotone{sulentic.fig4ab.new.ps}
\caption{UPPER: Interference filter images. LEFT (a):  Continuum-subtracted
line image centered at 6738$\rm{\AA}$ which includes H$\alpha$ and
[NII] emission from SQ. There is contamination from [NII] in the new
intruder. RIGHT (b):  Equivalent image centered at 6668$\rm{\AA}$ and
imaging H$\alpha$ emission in the new intruder. Contour maps for a) and
b)  with appropriate flux levels can be found in Xu et al. (1999).  The
nuclei of NGC7317 as well as NGC7318a and b are  marked with an
``X''.  Scale can be obtained from the 21 arcsec separation of the X's
for NGC7318a and b.
\label{haha}}
\end{figure}
\vfill\eject

\begin{figure}
%\plotone{sulentic.fig5ab.new.ps}
\caption{LOWER: HI 21cm radio contours superposed on above images. LEFT (c): HI
contours for velocities in the SQ range 6475-6755 km/s. RIGHT (d): HI
contours for velocities in the new intruder range (5597-6068 km/s). HI
contours levels correspond to 5.26e19, 1.58, 2.63, 5.26,
7.89$\times$10$^{20}$ cm$^{-2}$  with last contour shown only in Figure
4c. 
\label{hahi}} 
\end{figure}
\vfill\eject

\begin{figure}
%\plotone{sulentic.fig6.eps}
\caption{The lower part of the schematic shows the wavelength ranges for
H$\alpha$ and [NII]$\lambda$6583 emission in SQ and the new intruder as
well as [SII]$\lambda\lambda$6717,31 emission in foreground NGC7320.
The upper part shows the wavelength ranges sampled by various filters
and Fabry-Perot observations reported here and elsewhere. In the top
panel ``Plana et al.'' refer to published CFHT and Russian 6m  F-P
observations of Plana et al. (1999).  SPM IF and CA IF refer to
interference filter images obtained at San Pedro Martir and Calar Alto
respectively while SPM F-P refers to Fabry-Perot observations taken
with PUMA at San Pedro Martir.
\label{ranges}}
\end{figure}
\vfill\eject

\begin{figure}
%\plotone{sulentic.fig7.new.ps}
\caption{Identification charts (H$\alpha$ contours) for all emission regions
with measured velocities from our Fabry-Perot observations for (a) SQ
([NII]$\lambda$6583) and (b) the NI (H$\alpha$). Velocities for the
identified features are found in Table 2. 
\label{fpid}}
\end{figure}
\vfill\eject

\begin{figure}
%\plottwo{sulentic.fig8a.new.ps}{sulentic.fig8b.new.ps}
\caption{ISO images of SQ: a) 11.4$\mu$m MIR ISOCAM and b) 60$\mu$m
FIR ISOPHT. Images are in log scale. The display range for the 11.4$\mu m$ 
image is [0.01, 10] mJy/pix (6 arcsec square pixels). The 15 micron 
contours are 0.1, 0.2, 0.3, 0.4, 0.8, 1.1, 1.4 and 1.7 mJy on 
a 60$\mu m$ image with a display range [0.1, 10] mJy/pix.  
The bulk of the emission originates from the Seyfert 2 
nucleus of NGC7319 and the late-type foreground spiral NGC320.   
\label{isoim}}
\end{figure}
\vfill\eject

\begin{figure}
%\plottwo{sulentic.fig9a.new.ps}{sulentic.fig9b.new.ps}
\caption{a) CFHT-B and b) B-R images of SQ. The inferred size of the 
new intruder is indicated by an ellipse in b) which is
windowed to emphasize the bluest features (white). Using the same data Mendes de
Oliveira et al. 2001 measure B-R=1.32 for SQ starburst A and
B-R=0.4-0.7 for the blue intruder emission regions north of it. \label{cfht}}
\end{figure}
\vfill\eject

\begin{figure}
\epsscale{0.7}
%\plotone{sulentic.fig10.new.ps}
\caption{Vignettes of eight sources or regions in SQ taken from the average
B-band WfPC2 image obtained with Hubble Space Telescope. The finding
chart is an R-band image obtained with the CA 1.5m telescope (see Moles
et al. 1998). Vignettes show:  A) the region of ISO starburst B in the
younger tidal tail with associated dust lane; B) Very blue stellar
concentration at the tip of the younger tidal tail. C) Emission region
near the north edge of the stripped HI cloud; D) NE corner of NGC7319
showing a concentration of blue stars on the end of the ring as well as
a large (possibly superposed) dust patch; E) region of starburst A
(lower left) and superposed intruder emission regions and dust lanes; F)
region of the new intruder and the shock front between starburst A and
new intruder central bulge showing radial features (structural smearing
or spokes as in the Cartwheel); G) portion of the disrupted arm or ring
of new intruder with ablating HII regions; H) The center of the shock
with debris of ablated HII regions and an emission ring (S edge).
\label{montage}}
\end{figure}
\vfill\eject

\begin{table}
\tablenum{1}
\caption{Net counts, fluxes and luminosities for the different HRI components 
in Stephan's Quintet.  Fluxes and luminosities in the 0.1-2.0 keV range
are derived from the total count rates with a conversion factor of
3.9$\times 10^{-11}$ ergs cm$^{-2}$ s$^{-1}$ and are corrected for the
line-of-sight absorption.  The useable  integration time is t= 77.5 ks. }

\label{components}
\begin{flushleft}
\begin{tabular}{lrrrrrr}
\tableline
\tableline
\noalign{\smallskip}
Comp.& \multicolumn{3}{c}{Net counts} & Intrinsic X-ray Flux & X-ray Luminosity
 \\
&old&new&total& (erg cm$^{-2}$ s$^{-1}$)& (erg s$^{-1}$) \\
 \noalign{\smallskip}
\tableline
 \noalign{\smallskip}
total r$<3'$& 370$\pm$48 &833$\pm$78&1203$\pm$92 & 6.1$\times 10^{-13}$ & 5.3$
\times 10^{41}$  \\
total r$<1.5'$& 305$\pm$28&703$\pm$45& 1008$\pm$53& 5.1$\times 10^{-13}$ & 4.5$
\times 10^{41}$  \\
Sey & 45$\pm$8& 117$\pm$13& $162\pm15$& 8.2$\times 10^{-13}$  &7.2$\times 10^{4
0}$ \\
shock (int)&  66$\pm$9 &164$\pm$15& $230\pm18$& 1.2$\times 10^{-13}$ &
1.0$\times 10^{41}$ \\
shock (ext)& 121$\pm$15&338$\pm$25& $459\pm29$& 2.3$\times 10^{-13}$ &
2.0$\times 10^{41}$ \\
N7318a &24$\pm$6&37$\pm$8& $61\pm10$&3.1$\times 10^{-14}$  &3.7$\times 10^{40}$
 \\
NW ext. &11$\pm$5&28$\pm$8& $39\pm9$&2.0$\times 10^{-14}$  &2.4$\times 10^{40}$
 \\
residual r$<1.5'$& 105$\pm$22&183$\pm$34& 288$\pm$41& 1.4$\times
10^{-13}$&1.3$\times 10^{41}$ \\
 \noalign{\smallskip}
\tableline
\noalign{\smallskip}
       \end{tabular}
       \end{flushleft}
\end{table}
\vfill\eject

\begin{table}
\tablenum{2}
\caption{ISO MIR and FIR FLUXES}
\begin{flushleft}
\begin{tabular}{lrrl}
\tableline
\tableline
\noalign{\smallskip}
SOURCE & S$_{11.4\mu m}$ & S$_{60\mu m}$ & log L$_{60 \mu m}$ \\
ID & mJy & mJy & L$_{\odot}$ \\
\tableline
NGC7319 & 47.5 & 529 & 9.83 \\
NGC7320 & 59.8 & 487 & 7.92  \\
NGC7318ab & 34.5 & $>$24 & $>$8.34 \\
Starburst A & 14.1 & 79.5 & 8.98 \\
Starburst B & 0.3 & &  \\
\tableline
\noalign{\smallskip}
         \end{tabular}
         \end{flushleft}
\end{table}

\begin{table}
\tablenum{3}
\caption{FABRY-PEROT (and other) VELOCITIES}
\begin{flushleft}
\begin{tabular}{lrrl||lrl}
\tableline
\tableline
\noalign{\smallskip}
\multicolumn{4}{c||}{Intruder }&\multicolumn{3}{c}{Stephan's Quintet} \\
Region & F-P & Plana& Lit. & Region & F-P& Identif./Notes \\
\#&   \multicolumn{3}{c||}{vel. (km s$^{-1}$)}&\# & vel. (km s$^{-1}$)  \\
\tableline
1 &5636* &--   &  5550h &38&7078+ \\
2 &5635* &5580 &  5400h/5594f  &39&6800*&  \\
3 &5680* &5640 &  5700h/5623f &40&6627+&  \\
4 &5640+ &-- &&41&6617* &starburst B  \\
5 &5708* &5715 &  5860e &42&6690* &starburst A  \\
6 &5727* &5715 &  5717o/5540/6500h  &43&6725* &NGC7319 Seyfert nucleus  \\
7 &5766+ &5740 &  6460 &44&6422+&  \\
8 &5775* &5750 &  6460 &45&6350/5987+  \\
9 &5750* &-- &&46&6370-&  \\
10&5737* &-- &&47&6370-&  \\
11&5795* &5765 && 48&5990+&  \\
12&5731* &-- &&49&6000+  \\
13&5699* &-- &&50&6350/6056+  \\
14&5766* &-- &&51&800+& NGC7320 [SII]  \\
15&5770* &5730 &&& \\
16&5780* &5755 &&& \\
17&5660* &-- &&& \\
18&5810+ &5815 &&& \\
19&5830* &-- &&& \\
20&5890* &5855 &&& \\
21&5715+ &-- &&& \\
22&5860- &-- &&& \\
23&5959* &-- &&& \\
24&5921+ &5935 &&& \\
25&5940- &-- &&& \\
26&5979* &5950 & 6020 \\
27&5990* &5960 & 6020 \\
28&5990* &-- &&& \\
29&5988+ &5960 &&& \\
30&6017* &-- &&& \\
31&6010* &5985 &&& \\
32&6000- &-- &&& \\
33&6017- &6005 &&& \\
34&5987- &6000 &&& \\
35&5997- &--   &&& \\
36&6036- &--  &&& \\
37&5959+ &-- &&& \\
\tableline
\noalign{\smallskip}
         \end{tabular}
         \end{flushleft}
\end{table}

\end{document}